\documentclass[aps,prb,twocolumn,groupedaddress,showpacs]{revtex4}

\usepackage{amsmath}
\usepackage{graphicx}
\usepackage{color}

\begin{document}


\title{Theory of edge-state optical absorption\\ in two-dimensional transition metal dichalcogenide flakes}



\author{Maxim Trushin}
\affiliation{Department of Physics, University of Konstanz, D-78457 Konstanz, Germany}
\author{Edmund J. R. Kelleher}
\affiliation{Department of Physics, Imperial College London, London, SW7 2AZ, UK}
\author{Tawfique Hasan}
\affiliation{Cambridge Graphene Centre, University of Cambridge, Cambridge, CB3 0FA, UK}

\date{\today}

\begin{abstract}
We develop an analytical model to describe sub-bandgap optical absorption in two-dimensional semiconducting transition metal dichalcogenide (s-TMD) nanoflakes.
The material system represents an array of few-layer molybdenum disulfide crystals, randomly orientated in a polymer matrix. 
We propose that optical absorption involves direct transitions between electronic edge-states and bulk-bands, depends strongly on the carrier population, and is saturable with sufficient fluence.
For excitation energies above half the bandgap, the excess energy is absorbed by the edge-state electrons, elevating their effective temperature.
Our analytical expressions for the linear and nonlinear absorption could prove useful tools in the design of practical photonic devices based on s-TMDs. 
\end{abstract}


\maketitle

\section{Introduction}

\begin{figure}
\includegraphics[width=\columnwidth]{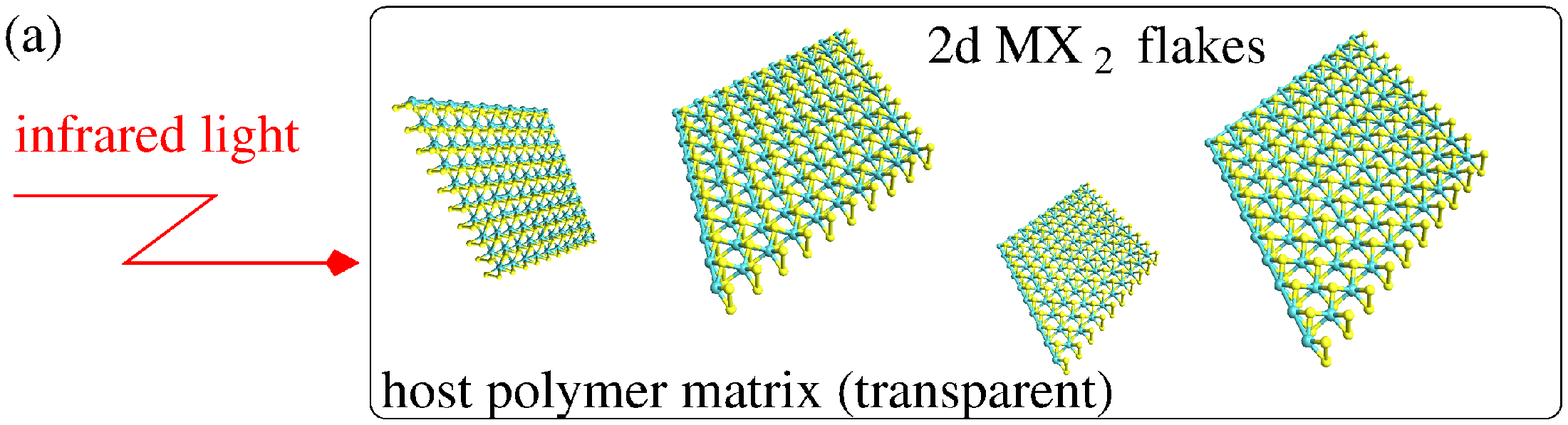}
\includegraphics[width=\columnwidth]{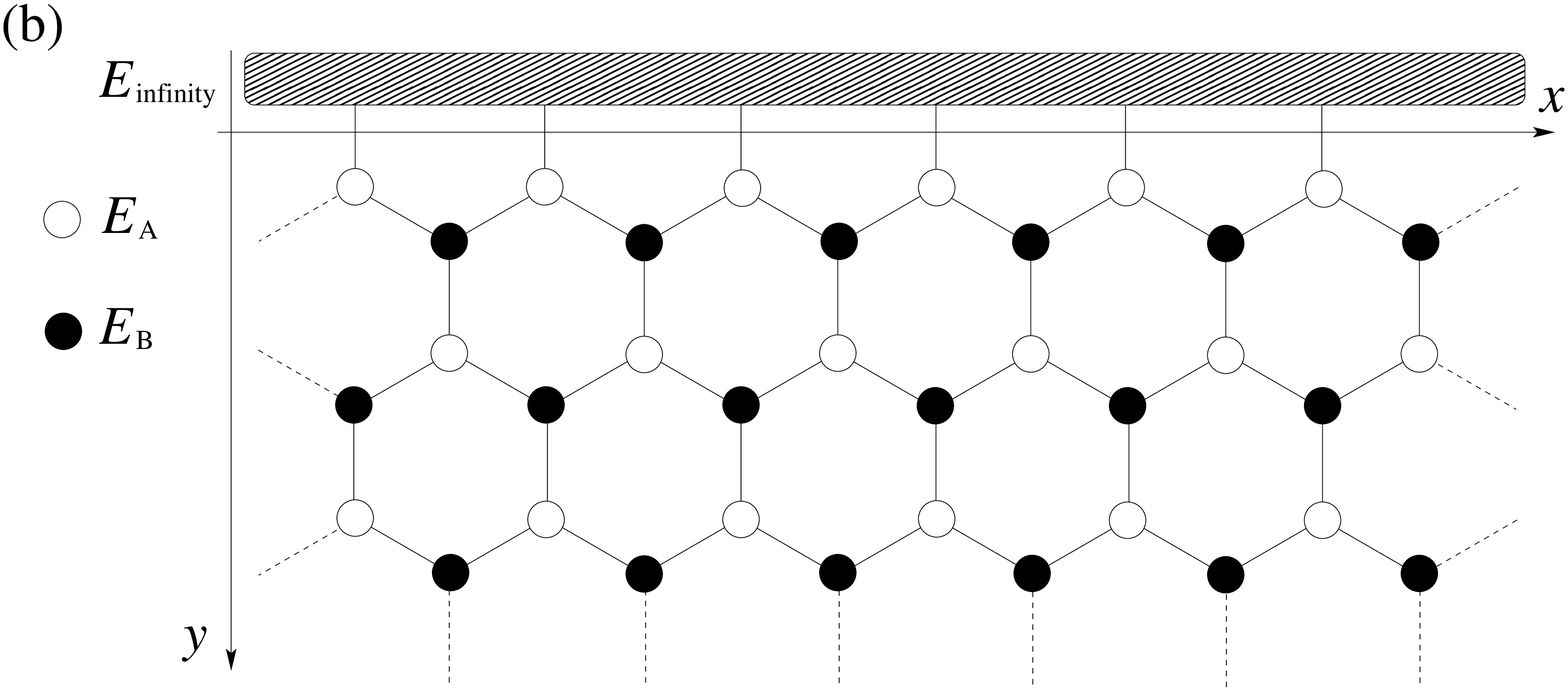}
\includegraphics[width=\columnwidth]{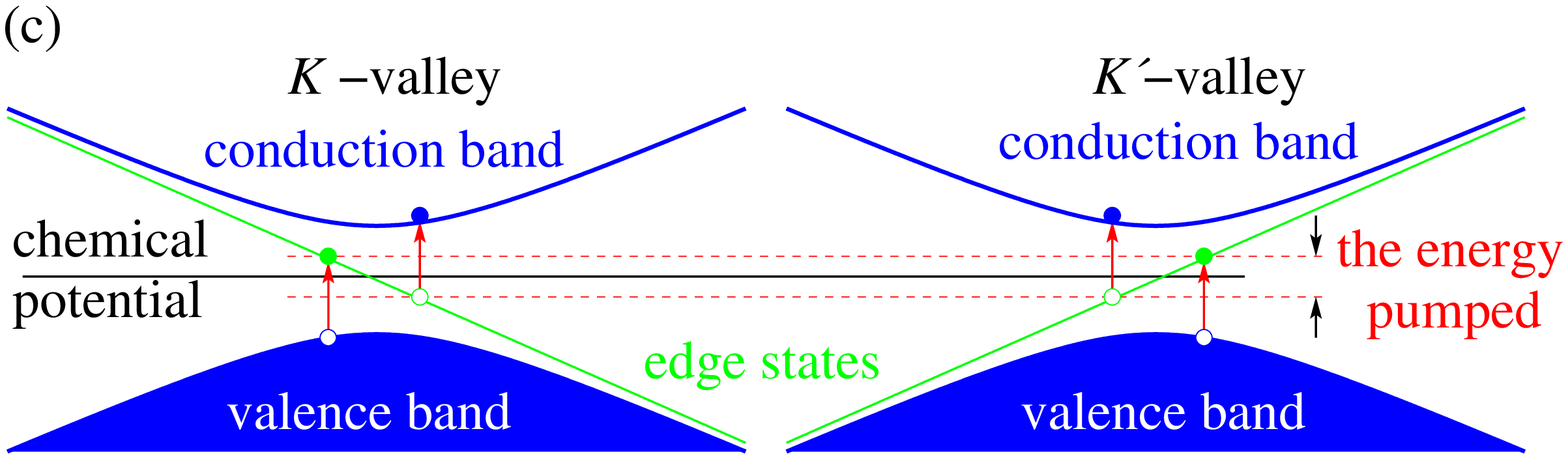}
\caption{\label{Fig1} (Color online.) (a) The s-TMD flakes are randomly distributed within a host polymer matrix illuminated by infrared light with an excitation energy below the material bandgap. (b) The honeycomb lattice with the on-site energies $E_A$ and $E_B$ terminated by the barrier along $x$ axis, where $E_B$ becomes infinite.
(c) The electronic band structure of a single flake includes conduction, valence and edge-states.
The edge-states are one-dimensional, i.e. the depicted momentum axis is parallel to the flake's edge.
There are two mirror copies of these bands in the first Brillouin zone (K and K' valley).
For a given excitation energy, two {\em independent} optical absorption channels are possible in each valley
corresponding to the valence-to-edge and edge-to-conduction bands direct transitions. These transitions are shown by red arrows, and the electrons and holes created are depicted by the filled and empty circles, respectively, see also Fig. 6 in Ref. \cite{Peterfalvi2015}
Each edge-state electron-hole pair accumulates a certain amount of energy which after thermalization appears as an elevated temperature for the edge-state electrons. }
\end{figure}

In the last decade, following the discovery of graphene,\cite{Geim2011} research of two-dimensional (2d) materials has experienced an explosive growth.
A 2d material represents an atomically thin solid flake, with optical properties qualitatively different from its three-dimensional (3d) parent crystal.\cite{Xu2013,Wang2012}
One of the largest families of 2d materials is the transition metal dichalcogenides (TMDs) that contains over 40 different forms, either metallic or semiconducting.\cite{Xu2013}
TMDs have the general formula MX$_2$, where M represents a transition metal, (e.g. molybdenum or tungsten), and X represents a chalcogen (e.g. sulfur, selenium, tellurium).\cite{Xu2013,Wang2012}
Single-layer MX$_2$ crystals are quasi-2d structures, containing a plane of metal (M) atoms covalently bonded between two planes of chalcogen (X) atoms, see Fig.~\ref{Fig1}a.
In contrast to bulk semiconducting TMD (s-TMD) crystals, their monolayers typically exhibit a direct bandgap at visible or near-infrared frequencies, making them a suitable material for a range of photonic and optoelectronic applications.\cite{Mak2010,Korn2011,Splendiani2010,Xu2013} 
In a direct bandgap semiconductor, with a pristine lattice and of infinite extent, photons with energies lower than the bandgap cannot excite direct interband transitions; thus, single-photon absorption at these energies does not occur. 
Recent experiments by several research groups, however, have demonstrated both non-negligible linear absorption at sub-bandgap photon energies, as well as a finite nonlinear optical response in a variety of s-TMDs, including MoS$_2$\cite{Zhang2015,Woodward2014}, WS$_2$\cite{Zhang2015b, Mao2015}, and MoSe$_2$.\cite{Woodward2015b,Luo2015}
Liquid phase exfoliated MoSe$_2$-polymer composites, for example, have been reported to exhibit $>$7\% linear absorption in the 0.65--0.8~eV range,\cite{Woodward2015b} in spite of MoSe$_{2}$ having a direct (in monolayer form) and indirect (bulk) bandgap of $\sim$1.5-1.58~eV and $\sim$1.1~eV, respectively.\cite{Zhang2014,Wang2012}

Several mechanisms have been proposed to explain this phenomenon.
Supported by first principle calculations, Wang \emph{et al.} suggested that a reduction in the MoS$_2$ bandgap could be achieved by introducing crystallographic defect states.\cite{Wang2014}
The authors also suggested that defects could activate the material as a broadband saturable absorber.\cite{Wang2014}
We recently proposed that edge-states contribute to sub-bandgap absorption in s-TMDs.\cite{Woodward2015,pssb2016howe}
This mechanism is supported by earlier photothermal deflection spectroscopy of MoS$_2$ nanoflakes, where increased linear absorption at sub-bandgap energies was observed for large MoS$_2$ crystals after lithographic texturing that increased the total amount of edges in the sample.\cite{Roxlo1987} 
s-TMD flakes prepared by liquid phase exfoliation (LPE) -- a widely used technique for the low-cost, mass manufacture of nanomaterials -- also have a high edge to surface area ratio, and are thus expected to exhibit sub-bandgap states, supporting absorption of photons with lower energies than the material bandgap.
Recent studies have demonstrated that the sub-bandgap absorption in s-TMD nanoflakes can be saturated, and exploited this effect in the development of ultrafast lasers operating in the near-infrared, corresponding to photon energies in the range 0.6--1.12 eV.\cite{Zhang2015,Woodward2014,Zhang2015b, Mao2015,Woodward2015b,Luo2015,Woodward2015}
While a growing body of experimental work continues to substantiate the process of sub-bandgap absorption in s-TMDs, and practical applications of this phenomenon are being leveraged in the field of photonics, theoretical analyses are limited and the origin of sub-bandgap optical absorption remains an open question. 
Here, we develop an analytical theory, testing the hypothesis of edge-mediated absorption in s-TMDs to explain the phenomenon of sub-bandgap saturable absorption.

The electronic states at the edges of a nanoflake (edge-states) have been modeled to date using two approaches.
Firstly, by focusing on the atomic structure of a particular edge and
computing the energy dispersion by means of a tight-binding Hamiltonian with appropriate boundary conditions
\cite{Brey2006,Akhmerov2008,Lado2015,Pavlovic2015,Peterfalvi2015,Segarra2015}; secondly, using density functional theory (DFT).\cite{PRL2001bollinger,Bollinger2003,JAMS2008li,Vojvodic2009,JPCL2012kou,Erdogan2012,PRB2014xu}
It has been shown a few years ago\cite{RSC2012pan} that the chalcogen-terminated zig-zag edges are the most stable because they have lowest energies
without hydrogen saturation. We therefore expect such edge types to be the most abundant in the dispersion of non-hydrogenated 2dTMDs.
Moreover, such edges maintain one-dimensional (1d) metallic states, as confirmed by ab-initio\cite{JAMS2008li,PRB2014xu}
and continuum-model\cite{Peterfalvi2015} calculations.
The latter shows that the band structure of the purely dichalcogen-terminated zig-zag edge can be well 
approximated by 1d bands with linear dispersions, where electrons are propagating in opposite directions in the K and K' valleys.
Our model shown in Fig. \ref{Fig1}(b,c) mimics this behavior, but, in contrast to the previous approches, 
allows us to calculate the wave functions and the Fermi's golden-rule optical transitions from and to the edge-states analytically.
In detail, we use an effective Hamiltonian proposed in Ref.\cite{Xiao2012} but with a spatially dependent bandgap simulating the flake edge.
A somewhat similar model is known in the literature as a neutrino billiard.\cite{Berry1987}

The ab-initio calculations reviewed above are able to provide a quantitative description of the optical absorption of a particular flake with a given edge type;
however, experimental measurements are typically preformed on an array of small flakes, randomly oriented in a polymer, with different edge types. 
We therefore need an effective model which focuses on the most optically active metallic states supported by the most stable chalcogen-terminated zig-zag edges. 
The model may not be valid for isolated flakes that may not possess metallic edge-states.
Nonetheless, it should provide a reliable optical absorption estimate for a large ensemble of flakes, where optically inert edge-states
are dominated by their active counterparts. 
Focusing on the most important edge type allows for explicite expressions for the linear and staturable optical absorptions.
The compromise for this simplification is the lack of predictive power on the quantitative level.

The peculiarities of the edge-state absorption are depicted in Fig.~\ref{Fig1}c.
In contrast to the two-band model for bulk semiconductors,\cite{Garmire2000}
our approach involves {\em three} electron subsystems. 
A one-dimensional edge-state electron subsystem always remains in the metallic regime with the Fermi energy determined by the bulk chemical potential.
In contrast, the conduction and valence bands are in the semiconducting regime:
the valence band is occupied almost completely whereas the conduction band is nearly empty.
Subgap direct transitions occur between the valence band and edge states as well as the edge and conduction band states. 
The relative contribution of these two transitions is determined by Pauli blocking and depends on the relationship between the excitation frequency and the Fermi level.
We show, that despite the complexity of the model, the saturable subgap absorption  $A^\Phi$ for s-TMD flakes
can be written in the conventional form \cite{Garmire2000}
\begin{equation}
\label{sat2}
 A^\Phi=\frac{A}{1+\frac{\Phi}{\Phi^s}}, 
\end{equation}
where $A$ is the relative linear absorption estimated by Eq. (\ref{ultimate}), $\Phi$ is the incident
fluence, $\Phi^s$ is the saturation fluence given by Eq. (\ref{ultimate2}).
The absorption is defined as a ratio of the absorbed radiation fluence to the incident fluence. 
In the rest of the paper, we derive the analytical expressions for $A$ and $\Phi^s$, and analyze their behavior.

\section{Model}
\label{mod}

From the point of view of the band theory, the difference between semiconductor and vacuum can be described by means of the bandgap $\Delta$: it is finite in the semiconducting region but infinite outside, where no conduction is possible. 
Let us consider a simple Hamiltonian derived for electrons on  a honey-comb lattice using the tight-binding approach 
with  the lattice constant $a$, the on-site energies $E_{A,B}$, and the nearest-neighbor hopping $t_\perp$. 
Near the K corner of the hexagonal first Brillouin zone, the Hamiltonian can be  written in the continuum limit as \cite{McCann2012}
$$
H_0^K=  \left( \begin{array}{cc}
    E_A & -t_\perp \frac{\sqrt{3} a}{2}(\hat{k}_x - i \hat{k}_y) \\ 
    -t_\perp \frac{\sqrt{3}a}{2}(\hat{k}_x + i \hat{k}_y) & E_B \\ 
  \end{array}\right),
$$ 
where $\hat k_x=-i \partial_x$, $\hat k_y=-i \partial_y$ are momentum operators.
(The Hamiltonian for K'-corner can be obtained by the substitution $\hat{k}_x\to -\hat{k}_x$.)
This Hamiltonian can be rewritten in a more instructive form  given by \cite{Xiao2012}
\begin{equation}
\label{ham1}
H_0^K= \mathrm{const} + \left( \begin{array}{cc}
    \frac{\Delta}{2} & \hbar v(\hat{k}_x - i \hat{k}_y) \\ 
    \hbar v(\hat{k}_x + i \hat{k}_y) & -\frac{\Delta}{2} \\ 
  \end{array}\right),
\end{equation}
where $\mathrm{const}=(E_A + E_B)/2$, $\Delta=E_A - E_B$ represents the bandgap,
and $- \sqrt{3}at_\perp/2=\hbar v$, with $\hbar v= 1.1~\mathrm{eV} \times 3.193$\AA\, for MoS$_2$.~\cite{Xiao2012}
The gap can be either positive or negative depending on the difference between the on-site energies $E_{A,B}$.
The spin-orbit coupling is neglected here. It results in the valley-spin locking which, in turn, can be used
for the valley-selective pump-probe spectroscopy with circularly polarized light. 
Since we are dealing with the linear polarization, both valleys contribute equally and the only effect
of the spin-orbit splitting is the spin-dependent bandgap.

The edge-states along the $x$-axis can be simulated by means of a $y$-dependent gap $\Delta(y)$.
We first solve the edge-state spectral problem for K-valley $H_0^K \psi_e = E_e \psi_e$ and obtain the eigen state wave function $\psi_e$ in the form 
\begin{equation}
\label{e-general}
\psi_e = C \exp\left(i k_x x - \int\limits_0^y \frac{ \Delta(y')dy'}{2 \hbar v}\right)\left( \begin{array}{c}
    1 \\ 
    -1 \\ 
  \end{array}  \right),
\end{equation}
where $C$ is a normalization constant, and $\Delta(y)$ should change its sign at $y=0$.\cite{Hasan2010}
An edge along the $y$-axis can be modeled in a similar way by an $x$-dependent gap $\Delta(x)$.
Since we aim for an analytical derivation of the linear absorption and saturation fluence, we simplify $\Delta(y)$ as
\begin{equation}
\label{gap}
 \Delta(y)= \left\{ \begin{array}{lll}
  \Delta>0, &  y \geq 0 & \quad \mathtt{(semiconductor)};\\ 
  -\infty, & y<0 & \quad \mathtt{(vacuum)}.\\ 
\end{array}\right.
\end{equation}

Eq.~(\ref{e-general}) then reads
\begin{equation}
\label{e}
\psi_e = \sqrt{\frac{\Delta}{2L \hbar v}} \exp\left(i k_x x - \frac{y \Delta}{2 \hbar v}\right)\left( \begin{array}{c}
    1 \\ 
    -1 \\ 
  \end{array}  \right), \quad y \geq 0
\end{equation}
which is normalized as 
$$
\lim\limits_{W \to \infty}\int\limits_0^{L} dx \int\limits_0^{W}dy ( \psi_e^\dagger \psi_e) = 1,
$$ 
and obeys the dispersion $E_e = -\hbar v k_x$. Due to Eq.~(\ref{gap}), $\psi_e$ exponentially vanishes in the bulk because $\Delta>0$ at $y\geq 0$.
Note that $\psi_e$ equals to zero at $y<0$ but is finite at $y=0$, i.e. it demonstrates a step-like behavior.
This is because $\Delta(y)$ is not a true electrostatic potential, as emphasized by Berry and Mondragon \cite{Berry1987}, but 
a ``staggered'' one.\cite{Akhmerov2008} The staggered potential depends on the sublattice, whereas  true electrostatic potential does not.
Even if $\Delta(y)$ goes to infinity, it is not equivalent to the hard-wall potential,
where the wave function must vanish at the border.  
For K'-valley, the solution of the spectral problem results in the same dispersion $E_e$ but taken with an opposite sign; see Fig.~\ref{Fig1}c.
In contrast to a topological quantum-Hall insulator,\cite{Hasan2010} the edge states (\ref{e}) exist in two mirror copies in two valleys.
To give an example, the edge-state electrons in MX$_2$ monolayers may experience intervalley backscattering,
i.e. the edge-state electron transport is not topologically protected.
It is worth emphasizing that our conclusions do not depend on whether the edge is along the $x$ or $y$ direction since
the optical absorption is averaged over the flake orientation.

The bulk conduction band eigen wave functions for K-valley are given by
\begin{equation}
\label{c}
\psi_c = \frac{1}{\sqrt{L W}} \exp\left(i k_x x + ik_y y\right) \left( \begin{array}{c}
    \cos\frac{\theta}{2} \\ 
    \sin\frac{\theta}{2}{\mathrm e}^{i \phi} \\ 
  \end{array}  \right),
\end{equation}
with the dispersion $E_c = \sqrt{(\hbar v k)^2 + \Delta^2 / 4}$, whereas the valence band wave functions read
\begin{equation}
\label{v}
\psi_v = \frac{1}{\sqrt{L W}} \exp\left(i k_x x + ik_y y\right) \left( \begin{array}{c}
    \sin\frac{\theta}{2} \\ 
    -\cos\frac{\theta}{2}{\mathrm e}^{i \phi} \\ 
  \end{array}  \right),
\end{equation}
with the dispersion  $E_v = -\sqrt{(\hbar v k)^2 + \Delta^2 / 4}$.
Here, 
$$
\tan \theta = \frac{2 \hbar v k}{\Delta}, \quad \tan \phi = \frac{k_y}{k_x}.
$$
The bulk states are normalized to unity on the rectangle $0\leq x \leq L$, $0\leq y \leq W$.

The electron-photon interaction Hamiltonian for K-valley 
is derived from (\ref{ham1}) and is given by\cite{Trushin2015,Trushin2011}
$$
H^\mathrm{int}=  \frac{e v E_0}{2 \omega} \left( \begin{array}{cc}
    0 & {\mathrm e}^{-i\theta_E} \\ 
    {\mathrm e}^{i\theta_E} & 0 \\ 
  \end{array}\right),
$$ 
where $E_0$, $\omega$, and $\theta_E$ are the electromagnetic wave amplitude, frequency, and polarization angle correspondingly.
The valence-to-edge states transitions are described by the following matrix element
\begin{eqnarray} \nonumber
&& \langle \psi_e \vert H^\mathrm{int} \vert  \psi_v \rangle = -\sqrt{\frac{\Delta}{2\hbar W v}} \frac{e v E_0}{2 L \omega}
\left(\cos\frac{\theta}{2}{\mathrm e}^{i\phi - i\theta_E}  \right.  \\
&&
\label{ev}
+ \left. \sin\frac{\theta}{2}  {\mathrm e}^{i\theta_E}\right) \frac{{\mathrm e}^{i(k_x - k'_x)L} - 1}{i (k_x - k'_x)} \frac{{\mathrm e}^{\left(ik_y - \frac{\Delta}{2\hbar v}\right)W} - 1}{i k_y - \frac{\Delta}{2\hbar v}}.
\end{eqnarray}
Here, $(k_x, k_y)=\mathbf{k}$ and $k'_x$ are momenta components in the bulk and at the edge, respectively.
The valence-to-edge states transition rate can be calculated as
\begin{eqnarray}
&& \label{ev2}
g^\mathrm{ph}_{ev} (\omega)= \sum\limits_{k_x,k_y,k'_x}\frac{2\pi}{\hbar}\vert  H^\mathrm{int}_{ev} \vert ^2 \left(f^{(0)}_v -  f^{(0)}_e \right) \\
&& \nonumber
\times \delta\left(-\hbar v k'_x + \sqrt{(\hbar v k)^2 + \Delta^2/4} - \hbar \omega \right),
\end{eqnarray}
where $f^{(0)}_v $,  $f^{(0)}_e$ are the Fermi-Dirac distributions for electrons in the valence band and 
in the edge-states, respectively and $\vert  H^\mathrm{int}_{ev} \vert ^2$ reads
$$
\vert  H^\mathrm{int}_{ev} \vert ^2 = \lim\limits_{L,W \to \infty} \vert   \langle \psi_e \vert H^\mathrm{int} \vert  \psi_v \rangle \vert^2
$$
$$
=\frac{\Delta}{2\hbar v} \frac{2 \pi}{L W} \delta(k_x - k'_x) \left(\frac{e v E_0}{2 \omega}\right)^2
\frac{1+ \sin\theta \cos(\phi - 2\theta_E)}{\left( \frac{\Delta}{2\hbar v} \right)^2 +k_y^2}.
$$
The edge-to-conduction band transition rate differs from Eq.(\ref{ev2}) by the sign in front of the $\theta_E$-dependent term and by the filling factors.
The corresponding generation rate reads
\begin{eqnarray}
&& \label{ce2}
g^\mathrm{ph}_{ce} (\omega)= \sum\limits_{k_x,k_y,k'_x}\frac{2\pi}{\hbar}\vert  H^\mathrm{int}_{ce} \vert ^2 \\
&& \nonumber
\times \delta\left( \sqrt{(\hbar v k)^2 + \Delta^2/4} + \hbar v k'_x  - \hbar \omega \right) \left(f^{(0)}_e -  f^{(0)}_c \right),
\end{eqnarray}
where
$$
\vert  H^\mathrm{int}_{ce} \vert ^2 = \lim\limits_{L,W \to \infty} \vert   \langle \psi_c \vert H^\mathrm{int} \vert  \psi_e \rangle \vert^2
$$
$$
=\frac{\Delta}{2\hbar v} \frac{2 \pi}{L W} \delta(k_x - k'_x) \left(\frac{e v E_0}{2 \omega}\right)^2
\frac{1- \sin\theta \cos(\phi - 2\theta_E)}{\left( \frac{\Delta}{2\hbar v} \right)^2 +k_y^2},
$$
and $f^{(0)}_c$ stands for the conduction band Fermi-Dirac distribution.


The flakes are randomly oriented, thus, the relative optical absorption is determined by the ratio between the $\theta_E$-averaged absorbed power
$\hbar \omega \langle g^\mathrm{ph}_{ev} + g^\mathrm{ph}_{ce} \rangle_{\theta_E}$ and the incident radiation power $(c E_0^2 S)/(8\pi)$ with $S$
being the illuminated area. To sum-up over $k'_x$, $k_x$, and $k_y$ we transform sums to integrals as
$$
\sum\limits_{k_x,k_y,k'_x} \to \int\frac{dk'_x L}{2\pi}\int\frac{dk_x L}{2\pi} \int \frac{dk_y W}{2\pi}.
$$
The integral over $k'_x$ is taken using the momentum conservation represented above as $\delta(k_x - k'_x)$. The integral over $k_x$ is then taken
using the energy conservation utilizing the transformation
$$
\delta\left( \sqrt{(\hbar v k)^2 + \Delta^2/4} \pm \hbar v k_x  - \hbar \omega \right) =
$$
$$
= \frac{\hbar^2 \omega^2 + \Delta^2/4 + \hbar^2 v^2 k_y^2}{2\hbar^3 \omega^2 v} \times
 $$
 $$
 \times \delta\left(k_x \mp \frac{\hbar^2 \omega^2 - \Delta^2/4 - \hbar^2 v^2 k_y^2}{2\hbar^2 \omega v} \right).
$$
We then substitute $\hbar v k_y=\varepsilon$, $E_\omega=\hbar\omega$ and obtain the relative absorption
of a single edge $A_1$ in the form $A_1=A_1^+ + A_1^-$, where
$A_1^\pm$ correspond to the  $v\to e$ and $e\to c$ transitions, respectively and are given by
\begin{equation}
 \label{main1}
 A_1^\pm=\frac{e^2}{\hbar c} \frac{\hbar v L}{ S} \frac{\Delta}{4E_\omega} \int\limits_{-\infty}^{\infty} d\varepsilon  
 \left(\frac{1}{E_\omega^2} + \frac{1}{\varepsilon^2  + \Delta^2/4} \right) F^\pm(\varepsilon).
\end{equation}
Here, $F^\pm(\varepsilon)$ describe the corresponding occupations and are given by
\begin{eqnarray}
&& \nonumber
F^+(\varepsilon)=
\frac{1}{1+\exp\left(-\frac{\varepsilon^2 +\Delta^2/4 + E_\omega^2}{2 E_\omega T_0} -\frac{\mu_p}{T_0} \right)} \\
&& \label{main2}
- \frac{1}{1+\exp\left(-\frac{\varepsilon^2 +\Delta^2/4 - E_\omega^2}{2E_\omega T} -\frac{\mu}{T} \right)},  \\
&& \nonumber 
F^-(\varepsilon)=\frac{1}{1+\exp\left(\frac{\varepsilon^2 +\Delta^2/4 - E_\omega^2}{2 E_\omega T} -\frac{\mu}{T} \right)}\\
&& \label{main3} 
-\frac{1}{1+\exp\left(\frac{\varepsilon^2 +\Delta^2/4 + E_\omega^2}{2 E_\omega T_0} -\frac{\mu_n}{T_0} \right)}.
\end{eqnarray}
Here, we set different (fluence dependent) quasi Fermi levels\cite{Nelson2004}  $\mu_n$ and $\mu_p$ for the conduction and valence bands correspondingly.
The quasi Fermi levels  $\mu_n$ and $\mu_p$ are both equal to the equilibrium chemical potential $\mu$ as long as no interband transitions occur
and no photocarriers are excited.
These notations will be utilized in section \ref{sat} devoted to the saturable absorption. 
Moreover, two temperatures have been introduced: $T_0$ is the lattice temperature for bulk electrons, and $T$ is the temperature
for edge-state electrons which may differ from $T_0$ in some cases described in section \ref{hot}.

We emphasize that Eq. (\ref{main1}) describes the optical absorption of a {\em single} edge of a {\em single} flake for
a {\em given} spin and valley channel. The total absorption of a s-TMD dispersion or a s-TMD-polymer composite should take into account different spin and valley channels
as well as the concentration of flakes. 
It can be shown that the K'-valley edge states result in the same contribution to the absorption as (\ref{main1}).
The spin-split absorption channels give two different contributions determined by the spin-dependent bandgap value $\Delta=\Delta_s$, but we neglect the spin splitting for the sake of simplicity.
Moreover, we assume that the flakes are squares of the size $d$, and all flakes are placed perpendicular to the light beam. 
To sum up these contributions, we define an {\em effective} length as
\begin{equation}
 \label{Leff}
 L^\mathrm{eff}= \ell S \quad \mathrm{with} \quad  \ell= 4d g_{sv} n_{2D},
\end{equation}
where $4d$ is the average perimeter of a flake, $g_{sv}=4$ is the spin/valley degeneracy, and
$n_{2D}$ is the number of monolayer flakes per unit area of a composite film.
The quantity $\ell$ then plays a role of the total effective length of monolayer flakes' edges per unit area of a composite film.
Assuming the size of the flake to be of the order of $100$nm, the monolayer flake concentration $n_{2D}\sim 10^{11}\,\mathrm{cm}^{-2}$
we estimate the effective length to be of the order of $1$km for a $1$mm$^2$ spot size.
In order to convert the absorption of a single edge (\ref{main1}) to the total absorption of a composite
we make the substitution $L\to L^\mathrm{eff}$, i.e. $A=A_1(L\to L^\mathrm{eff})$.
Eq. (\ref{main1}) is the main result of this work.
It can be used to calculate the linear and nonlinear absorption.
We now elaborate on these two cases.

\section{Linear absorption}
\label{lin}

\begin{figure}
\includegraphics[width=\columnwidth]{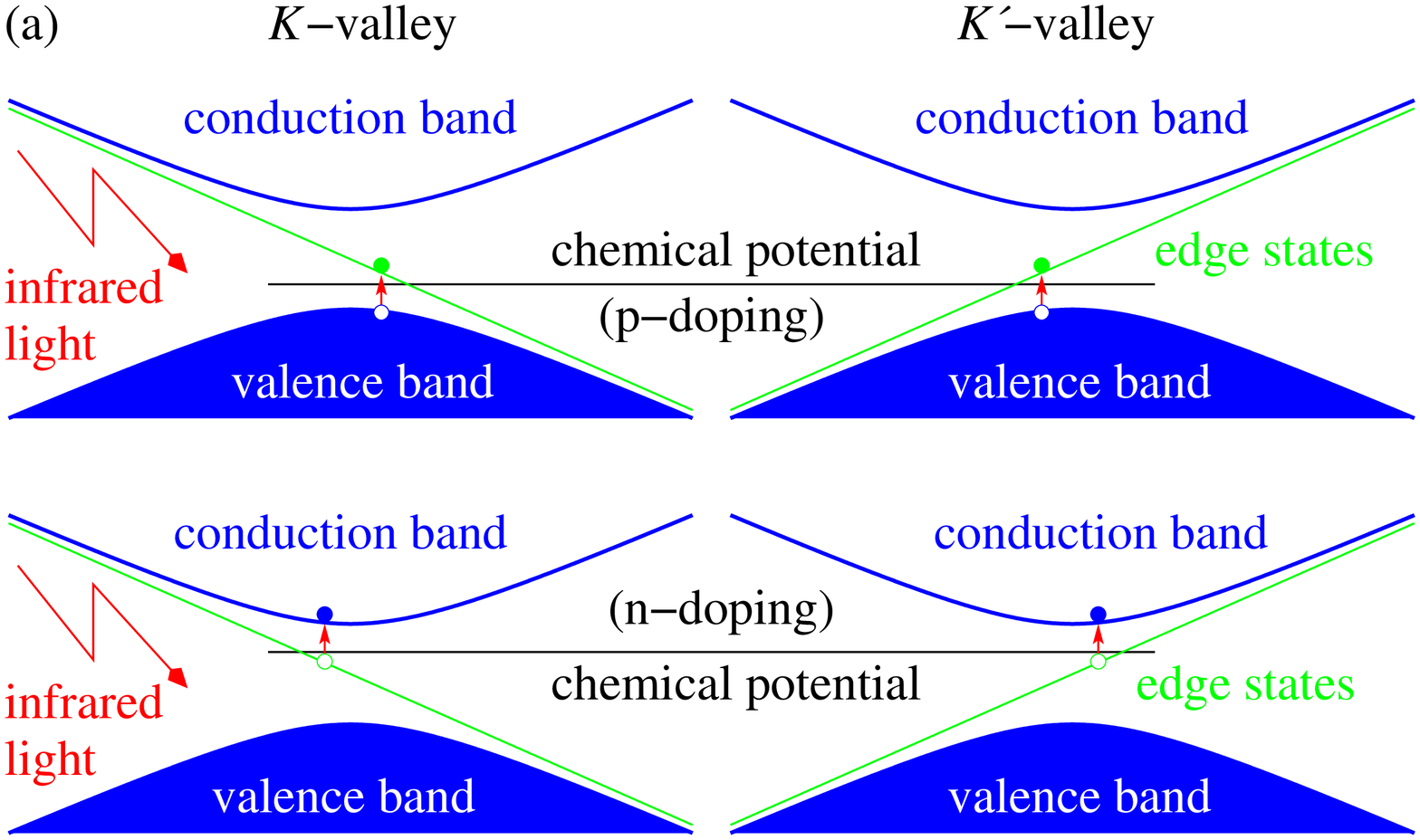}
\includegraphics[width=\columnwidth]{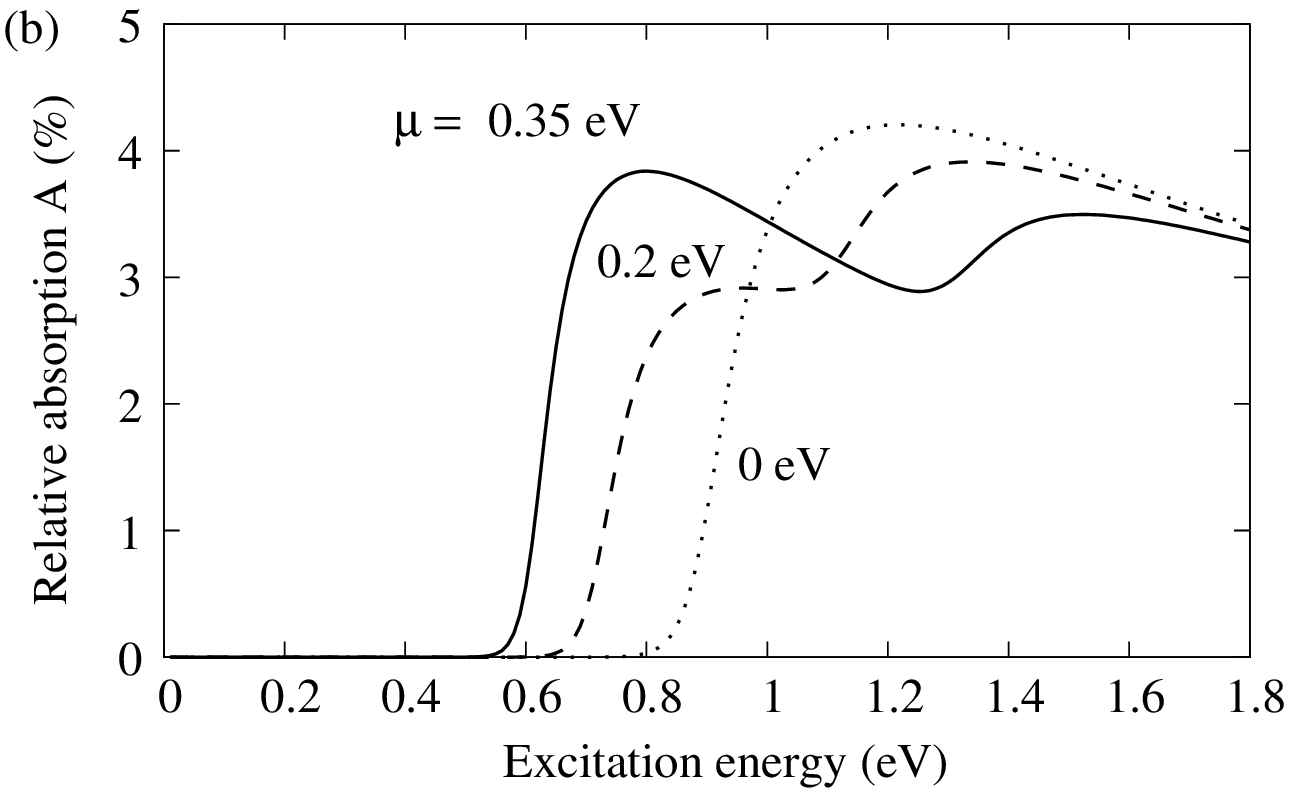}
\caption{\label{Fig2} (a) (Color online.) The possible direct optical transitions at a given radiation wavelength in doped samples.
Since the bands are symmetric there is no difference, whether the flakes are n- or p-doped.
(b) Relative linear optical absorption of a MoS$_2$ dispersion at room temperature computed from Eq.~(\ref{main-linear}).
The average flake size $d=90$ nm and the monolayer flake concentration  $n_{2D}=5.64\cdot 10^{11}$ cm$^{-2}$ 
have been deduced from Refs. \cite{Woodward2015,pssb2016howe}
The bandgap $\Delta=1.8$ eV and the band parameter $\hbar v= 1.1 \mathrm{eV} \times 3.193$\AA\,  are taken from Refs. \cite{Xiao2012,Kim2016} 
The excitonic and direct valence-to-conduction interband transitions relevant at the excitation energies near $\Delta$ are not taken into account.
The flakes are n-doped with the chemical potential ranging from $0$ to $0.35$ eV.
The spin-orbit splitting is neglected.}
\end{figure}

In the low-fluence limit we set the valence band occupation to $1$ (completely filled) and the conduction band occupation to $0$ (completely empty).
Eq.(\ref{main1}) can be then written as 
\begin{equation}
 \label{main-linear}
 A_1^\pm(T)=\frac{e^2}{\hbar c}  \frac{\hbar v L}{S}\frac{\Delta}{4E_\omega}  \int\limits_{-\infty}^{\infty} d\varepsilon  \left(\frac{1}{E_\omega^2} + \frac{1}{\varepsilon^2  + \Delta^2/4} \right)
 $$
 $$
\times\frac{1}{1+\exp\left(\frac{\varepsilon^2 +\Delta^2/4 - E_\omega^2}{2E_\omega T} \pm \frac{\mu}{T} \right)}.
 \end{equation}
In the intrinsic semiconductor limit ($\mu=0$) both terms $A^\pm_1$ are the same.
In the limit of $T=0$ Eq.~(\ref{main-linear}) takes the form
\begin{eqnarray}
&& \nonumber
A^\pm_1(0)=\frac{e^2}{\hbar c} \frac{\hbar v L}{S} \frac{\Delta}{2E_\omega}
\left[\frac{\sqrt{E_\omega^2 \pm 2\mu E_\omega -\Delta^2/4}}{E_\omega^2} 
 \right. \\
&& \left. 
+ \frac{2}{\Delta}\arctan\left(\frac{\sqrt{E_\omega^2 \pm 2\mu E_\omega-\Delta^2/4}}{\Delta/2} \right)\right].
\label{T0}
\end{eqnarray}
Eq.~(\ref{T0}) is applicable only when the square roots are real, the corresponding terms should be set to zero otherwise.
Physically, vanishing absorption corresponds to the Pauli blocking depicted in Fig. \ref{Fig2}a.

The total linear absorption of a composite can be obtained by making the substitution $L\to L^\mathrm{eff}$
and is shown in Fig. \ref{Fig2}b as a function of the excitation energy.
To be specific, we consider the n-doped samples ($\mu>0$). The opposite case of $\mu<0$ results in 
the same behavior since the bands are assumed to be symmetric with respect to $E=0$ (the middle of the bandgap).
At too low excitation energies (when  $E_\omega^2 + 2\mu E_\omega -\Delta^2/4<0$) the absorption vanishes.
Increasing the excitation energy, we first activate the transitions from the edge states to the conduction band.
This results in the relative absorption of about 4\% for the  s-TMD--composite we consider.
The absorption decreases slightly with the excitation wavelength until the transitions from
the valence band to the edge states becomes activated at $E_\omega^2 - 2\mu E_\omega -\Delta^2/4>0$. 
The dependence $A(E_\omega)$ is therefore non-monotonic due to the different absorption channels
opened at different $E_\omega$. 
Note that the bands in real MX$_2$ samples are spin-split; therefore, we expect each of two maxima in $A(E_\omega)$
to split into two that results in a somewhat more complicated pattern.
At low doping ($\mu \to 0$) the two maxima merge into a single absorption maximum that can also be seen in Fig.~\ref{Fig2}. 

In order to estimate the absorption maximum by the order of magnitude we consider Eq.~(\ref{main-linear}) 
in the limit $\mu=0$ and $T=0$. The function has a maximum at $E_\omega=0.67\, \Delta$. At this excitation energy 
the total linear absorption of a composite film can be estimated as
\begin{equation}
\label{ultimate}
 A \sim \frac{4e^2}{\hbar c} \frac{\hbar v }{\Delta}\ell,
\end{equation}
where $\ell$ is defined in (\ref{Leff}). The physical meaning is clear: the absorption is larger for smaller $\Delta$ because the real-space width of the edge state (\ref{e}) is larger for smaller gaps. 
The absorption is proportional to the total length of edges $\ell$ (per unit square) involved in the absorption.
Substituting parameters relevant for MoS$_2$,\cite{Kim2016}
and using $d\approx 100$nm and $n_{2D}\approx 5\cdot 10^{11}\,\mathrm{cm}^{-2}$,\cite{Woodward2015,pssb2016howe} we obtain the subgap absorption of the order of 1\%.

\section{Saturable absorption}
\label{sat}

If the incident fluence $\Phi$ is close to the saturation fluence, then the quasi Fermi energies
$\mu_n$ and $\mu_p$ should be taken into account.
They can be calculated using the particle conservation.
On the one hand, the photocarrier concentration in the conduction band due to the single-edge absorption is $n^\mathrm{ph}=\Phi A_1^- /E_\omega$,
where $A_1^-$ is the edge-to-conduction band absorption, see Eq.~(\ref{main1}).
On the other hand, the same concentration can be calculated for the thermalized electrons as
$$
n^\mathrm{ph}=\int \frac{d^2 k}{4\pi^2} \frac{1}{1+ \exp\left(\frac{\sqrt{(\hbar v k)^2 + \Delta^2 / 4}-\mu_n}{T_0}\right)}
$$
\begin{equation}
\label{nph}
 \approx \frac{T_0 \Delta}{4\pi\hbar^2 v^2} {\mathrm e}^{\frac{\mu_n-\Delta/2}{T_0}},
\end{equation}
This approximation is valid as long as $(\Delta/2-\mu_n)/T_0\gg 1$.
Thus, $\mu_n$ can be determined from
\begin{equation}
\label{mun}
{\mathrm e}^{\frac{\mu_n}{T_0}}=\frac{4\pi\hbar^2 v^2}{T_0 \Delta}\frac{\Phi A_1^-}{E_\omega} {\mathrm e}^{\frac{\Delta}{2T_0}}.
\end{equation}
The quasi Fermi energy for the valence band $\mu_p$ is calculated 
in the same way using the photoexcited hole concentration
$p^\mathrm{ph}= \Phi A_1^+/E_\omega$ and its thermalized version,
which reads
$$
p^\mathrm{ph}=\int \frac{d^2 k}{4\pi^2}
\left(1- \frac{1}{1+ \exp\left(\frac{-\sqrt{(\hbar v k)^2 + \Delta^2 / 4}-\mu_p}{T_0}\right)} \right)
$$
\begin{equation}
 \label{pph}
 \approx \frac{T_0 \Delta}{4\pi\hbar^2 v^2} {\mathrm e}^{-\frac{\mu_p+\Delta/2}{T_0}}.
\end{equation}
Note, that $\mu_p < 0$. Hence, $\mu_p$ can be found from
\begin{equation}
\label{mup}
{\mathrm e}^{-\frac{\mu_p}{T_0}}=\frac{4\pi\hbar^2 v^2}{T_0 \Delta}\frac{\Phi A_1^+}{E_\omega} {\mathrm e}^{\frac{\Delta}{2T_0}}.
\end{equation}

\begin{figure}
\includegraphics[width=\columnwidth]{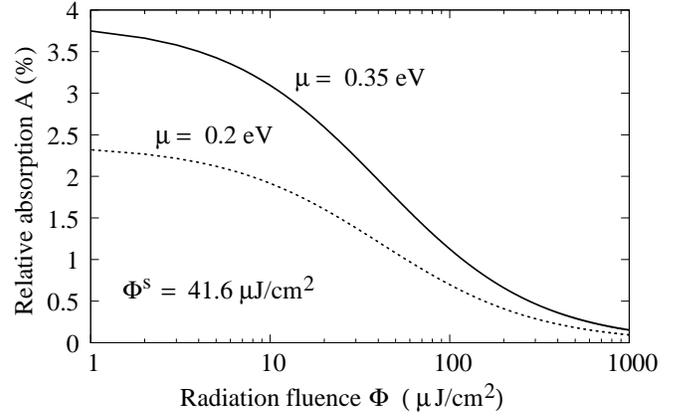}
\caption{\label{Fig3} Saturable optical absorption of a MoS$_2$ composite at $E_\omega=0.8$ eV. The parameters are the same as in Fig. \ref{Fig2}.
The saturation fluence is estimated from Eq.~(\ref{Phis}) with $L\to L^\mathrm{eff}$,
but Eq.~(\ref{ultimate2}) gives nearly the same result for $\Phi^s$ of about $40\,\mathrm{\mu J/cm^2}$.
It corresponds to the intensity of a few $\mathrm{MW/cm^2}$ at the electron-hole recombination time of the order of 10 ps, see Ref.\cite{Wang2015}
}
\end{figure}

Now, we employ Eqs.~(\ref{main2},\ref{main3}) assuming that 
$$
\frac{1}{1+\exp\left(-\frac{\varepsilon^2 +\Delta^2/4 + E_\omega^2}{2 E_\omega T_0} -\frac{\mu_p}{T_0} \right)}
$$
$$
\approx 1- \exp\left(-\frac{\varepsilon^2 +\Delta^2/4 + E_\omega^2}{2 E_\omega T_0} -\frac{\mu_p}{T_0} \right),
$$
$$
\frac{1}{1+\exp\left(\frac{\varepsilon^2 +\Delta^2/4 + E_\omega^2}{2 E_\omega T_0} -\frac{\mu_n}{T_0} \right)}
$$
$$
\approx \exp\left(-\frac{\varepsilon^2 +\Delta^2/4 + E_\omega^2}{2 E_\omega T_0} +\frac{\mu_n}{T_0} \right),
$$
and exclude $\mu_{n,p}$ using Eqs. (\ref{mun},\ref{mup}).
These approximations are standard for semiconductors:
we substitute the electron and hole Fermi-Dirac occupations by 
the corresponding Boltzmann distributions.
Note that the edge states are in the metallic regime and therefore, the Fermi-Dirac distribution must be retained for this subsystem.
To take the integral over $\varepsilon$, we calculate the following expressions:
$$
\int\limits_{-\infty}^\infty \frac{d\varepsilon}{E_\omega^2}
\exp\left(-\frac{\varepsilon^2 +\Delta^2/4 + E_\omega^2}{2 E_\omega T_0} +\frac{\Delta}{2T_0} \right)
$$
$$
=\frac{1}{E_\omega}\sqrt{\frac{2\pi T_0}{E_\omega}} 
\exp\left[-\frac{(E_\omega -\Delta/2)^2}{2 E_\omega T_0}\right],
$$
$$
\int\limits_{-\infty}^\infty \frac{d\varepsilon}{\varepsilon^2 + \Delta^2/4}
\exp\left(-\frac{\varepsilon^2 +\Delta^2/4 + E_\omega^2}{2 E_\omega T_0} +\frac{\Delta}{2T_0} \right)
$$
$$
=\frac{2\pi}{\Delta} 
\exp\left(\frac{\Delta - E_\omega}{2 T_0}\right) \mathrm{Erfc}
\left(\frac{\Delta}{\sqrt{8E_\omega T_0}} \right),
$$
where $\mathrm{Erfc}$ is the complementary error function.
After some algebra we obtain the saturable absorption in the form
\begin{equation}
 \label{sat1}
 A_1^\Phi=\frac{A_1(T)}{1+\frac{\Phi}{\Phi_1^s}},
\end{equation}
where $A_1(T)=A_1^+ + A_1^-$ is the linear absorption with $A_1^\pm$ given by Eq. (\ref{main-linear}). 
The saturation fluence $\Phi_1^s$ can be found from
\begin{equation}
\label{Phis}
\frac{1}{\Phi_1^s}=
\frac{\pi e^2}{\hbar c}
 \frac{\hbar^3 v^3 L }{E_\omega^3 T_0 S}
\exp\left[-\frac{(E_\omega -\Delta/2)^2}{2 E_\omega T_0}\right]
$$
$$
\times \left[\sqrt{\frac{2\pi T_0}{E_\omega}}
+\frac{2\pi E_\omega}{\Delta}\mathrm{e}^\frac{\Delta^2}{8 E_\omega T_0}
\mathrm{Erfc}
\left(\frac{\Delta}{\sqrt{8E_\omega T_0}} \right)
\right].
\end{equation}
If we neglect the heating of the edge state electrons, then we can set $T=T_0$ in Eq. (\ref{main-linear}),
and $A_1^\pm(T)$ can be approximated by $A_1^\pm(0)$ given by Eq. (\ref{T0}).
The nonlinear absorption $A_1^\Phi$ will be then determined solely by the $(1+\Phi/\Phi_1^s)^{-1}$ multiplier, as if
it is the standard two-band model.\cite{Garmire2000}
In order to find the total composite absorption we make the substitution $L\to L^\mathrm{eff}$ in (\ref{sat1}) and obtain our main result
(\ref{sat2}) with $\Phi^s=\Phi_1^s(L\to L^\mathrm{eff})$
and $A=A_1(L\to L^\mathrm{eff})$.

We show the composite nonlinear absorption $A^\Phi$ in Fig.~\ref{Fig3} at the telecommunication wavelength of $1550$ nm ($E_\omega=0.8$ eV).
The incident fluence can be translated to the intensity as $I=\Phi/\tau$ with $\tau$ being the electron-hole recombination time
of about 10 ps, see Ref.\cite{Wang2015}  The saturation fluence evaluated from (\ref{Phis}) in the excitation energy range 0.8--1.0~eV is of the order of 
$10\,\mathrm{\mu J/cm^2}$ that corresponds to the intensity of the order of $10^6 \,\mathrm{J /(s\cdot cm^2)}$,
relevant for the typical measurements.\cite{Woodward2015b}
Eq.~(\ref{Phis}) also suggests that the saturation intensity increases dramatically at the excitation energies far from $\Delta/2$.
Physically, the half of the bandgap $\Delta/2$ plays the same role in our approach
as the true bandgap $\Delta$ in the conventional two-band model.\cite{Garmire2000}
The saturation is most efficient when the photocarriers
are excited from and to the band edges. Our model is entering into this regime when the excitation energy is near $\Delta/2$, as one can see from Fig.~\ref{Fig1}c.
At the excitation energies much higher than $\Delta/2$,
the photocarriers are excited far from the conduction and valence band
edges and cannot be described by a thermalized distributions (\ref{mun}) and (\ref{mup}).
It is instructive to consider the limit $E_\omega=\Delta/2$ and
subsequently assume that $\Delta \gg T_0$.
The second term in Eq.~(\ref{Phis}) can be then approximated as
$\pi\mathrm{e}^\frac{\Delta}{4T_0}
\mathrm{Erfc}\left(\sqrt{\frac{\Delta}{4T_0}} \right)\approx
\sqrt{\frac{4\pi T_0}{\Delta}}$, and the final formula
for the composite saturation fluence reads
\begin{equation}
\label{ultimate2}
\frac{1}{\Phi^s} \sim \frac{4 \pi^{\frac{3}{2}} e^2}{\hbar c}
\frac{\hbar^3 v^3 \ell}{E_\omega^3 \, \sqrt{\Delta T_0}},
\end{equation}
where $E_\omega\sim \Delta/2$.
The absorption is therefore easier to saturate at smaller gap $\Delta$
and longer effective edge length defined in (\ref{Leff}).

\section{Hot electrons on edges}
\label{hot}

The situation becomes more complicated at the excitation energies higher than half the bandgap ($E_\omega > \Delta/2$) in the intrinsic semiconductor regime ($\mu=0$).
The energy necessary to promote one edge-state electron to the conduction band (or an edge-state hole to the valence band)
is $\Delta/2$. The question we address in this section is what happens with the excess energy $E_\omega - \Delta/2$ after each excitation event.

As already shown in  Fig.~\ref{Fig1}c, two independent excitation channels
corresponding to the valence-to-edge and edge-to-conduction band transitions are opened.
The valence-to-edge state transitions promote electrons to
just above the Fermi level at the same rate as the edge-to-conduction state transitions
create holes just below the Fermi energy, see Fig.~\ref{Fig1}c.
Effectively, these transitions lift an electron from an edge state below $\mu$ to another edge state above $\mu$. 
If the radiation intensity is high enough (the excitation is faster than the interband recombination), then this results in the generation of electron-hole pairs 
{\em within the edge-state subsystem}. Since electron-electron collisions are very efficient in a one-dimensional case the edge-state electron occupation quickly thermalizes 
to a Fermi-Dirac distribution with an elevated temperature.
Thus, the excess energy is accumulated by the edge-state electrons. Let us quantify this mechanism.

\begin{figure}
\includegraphics[width=\columnwidth]{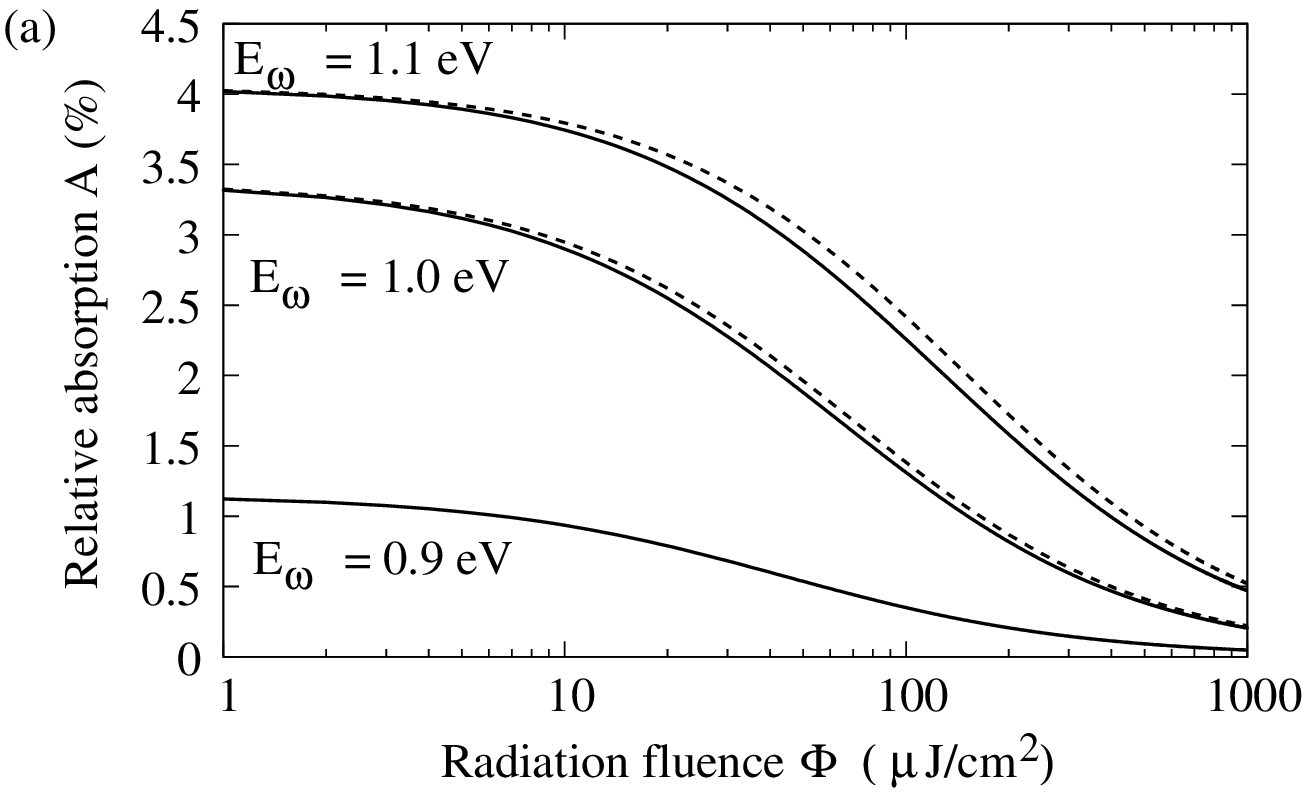}
\includegraphics[width=\columnwidth]{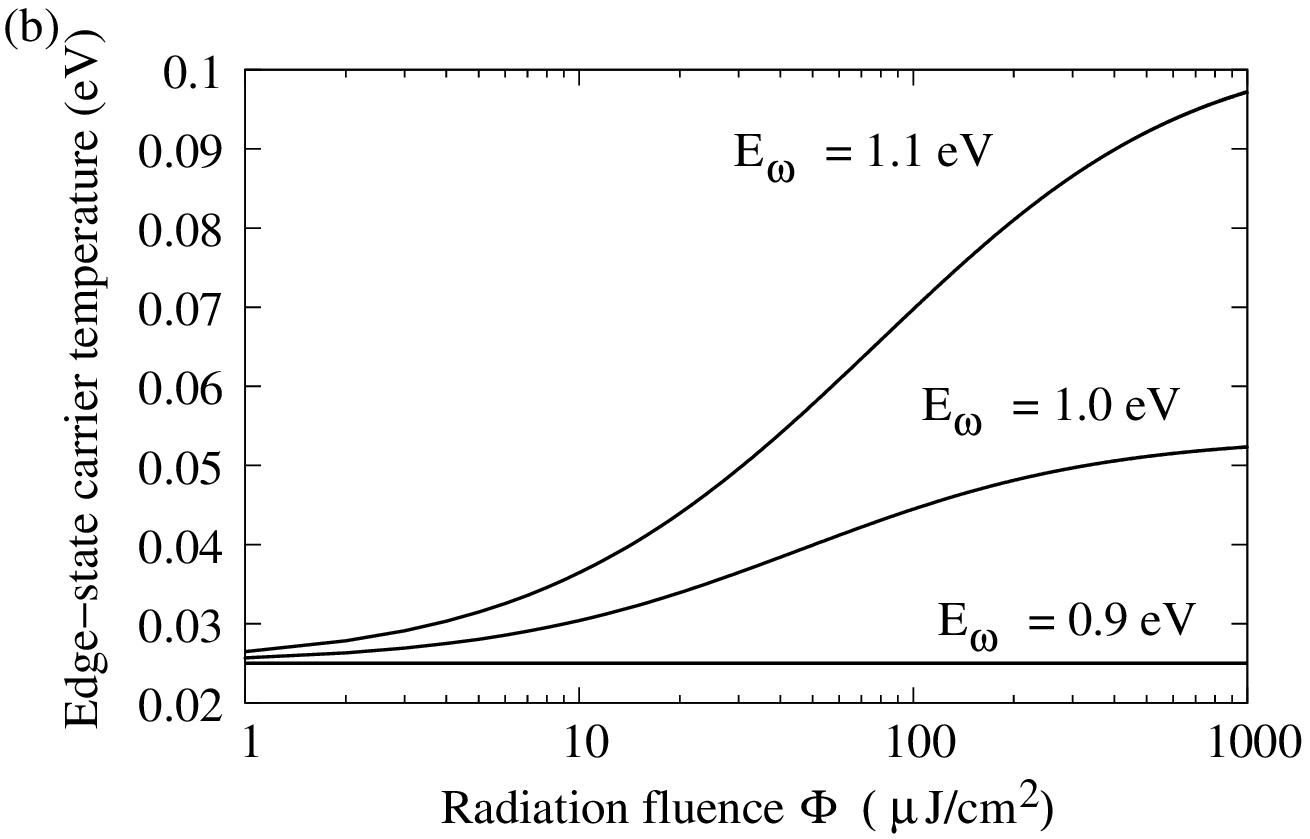}
\caption{\label{Fig4} (a) Saturable absorption in the intrinsic limit ($\mu=0$) at different excitation energies $E_\omega \geq \Delta/2$.
The dashed curves correspond to the simplified model where the edge-state electron temperature 
remains constant. The solid curves take into account the energy pumping due to the processes shown in Fig.~\ref{Fig1}c.
At the excitation energies higher than $\Delta/2$, the deviation between the solid and dashed curves is clearly visible.
(b) Edge-state electron temperature {\em vs.} fluence computed from (\ref{bal3}).}
\end{figure}

To calculate this temperature, we have to solve the energy balance equation
with respect to $T$:
\begin{equation}
\label{bal1}
\delta E + E_n + E_p = A_1 S \Phi.
\end{equation}
The right-hand side of (\ref{bal1}) is the absorbed  energy which is balanced with
the energies $\delta E$,  $E_n$, $E_p$ accumulated by
the thermalized edge, conduction and valence band electrons.
The last two can be estimated as $E_n\approx A_1^- S\Phi \Delta/2E_\omega$ and
$E_p\approx A_1^+ S \Phi \Delta/2E_\omega$, where the same approximation as in eqs. (\ref{nph}), (\ref{pph}) has been utilized.
Physically, $E_n$ ($E_p$) is the product between the photoexcited electron (hole) number  $A_1^\mp \Phi S /E_\omega$ and 
the typical energy $\pm\Delta/2$ in thermalized limit.
The energy pumped into the edge-state electron gas can be calculated assuming
that the occupation is already thermalized and given by the Fermi-Dirac distribution function.
The edge-state electron-hole excitation energy is the difference between 
the electron and hole energies within the edge-state band, as shown in Fig.~\ref{Fig1}c by dashed lines.
It can be written as
\begin{eqnarray}
 \nonumber
 E(T) && =  \int\limits_{-\infty}^{-\frac{\mu}{\hbar v}} \frac{dk_x L}{2\pi}
\frac{-\hbar v k_x}{1+\exp\left(\frac{-\hbar v k_x  - \mu}{T} \right)} \\
&& \nonumber 
 - \int\limits_{-\frac{\mu}{\hbar v}}^{\infty} \frac{dk_x L}{2\pi}
\frac{-\hbar v k_x}{1+\exp\left(\frac{\hbar v k_x  + \mu}{T} \right)} \\ 
&&=\frac{\pi T^2 L}{12\hbar v}.
\label{dE}
\end{eqnarray}
Note that $E(T)$ does not depend on $\mu$ because of the linear dispersion, hence, $\delta E$ is independent of $\mu$ as well and reads
\begin{equation}
\delta E = \frac{\pi L }{12 \hbar v}(T^2- T_0^2).
\end{equation}
Eq. (\ref{bal1}) is then written as
\begin{equation}
 \label{bal2}
 \frac{\pi (T^2- T_0^2)}{12\hbar v }\frac{L}{S} = \left(1-\frac{\Delta}{2 E_\omega}\right) \Phi A_1.
\end{equation}
Substituting $L\to L^\mathrm{eff}$, $A_1\to A^\Phi$ we obtain the following equation for the edge-state electron temperature in a composite:
\begin{equation}
 \label{bal3}
 T^2- T_0^2 = \frac{3\hbar e^2 v^2 \Phi \Delta }{\pi c E_\omega} \frac{1- \Delta/(2E_\omega)}{1+\Phi/\Phi^s}
 $$
 $$
 \times \sum_\pm  \int\limits_{-\infty}^{\infty} d\varepsilon 
 \frac{\left(E_\omega\right)^{-2} +\left(\varepsilon^2  + \Delta^2/4\right)^{-1}}{1+\exp\left(\frac{\varepsilon^2 +\Delta^2/4 - E_\omega^2}{2E_\omega T} \pm \frac{\mu}{T} \right)}.
\end{equation}
This equation can be solved with respect to $T$ numerically using the method of iterations (the method of consecutive approximations).
The result is demonstrated in Fig. \ref{Fig4}, where absorption and edge-state carrier temperature are shown for different
excitation energies. If $E_\omega=\Delta/2$, then the solid and dashed curves coincide, and heating of the edge-state electrons (solid curve) can be neglected.
If $E_\omega>\Delta/2$, then the excess energy $E_\omega - \Delta/2$ is pumped into the edge-state electron subsystem and
its temperature can reach $0.1$~eV ($\sim 1200$~K).
The high temperature makes the edge-states evenly populated in $k_x$-space that results
in less electrons excited from the edge-states to the conduction band and less empty space available for the 
electrons coming from the valence band. Hence, the elevated temperature slightly reduces absorption, as shown in Fig.~\ref{Fig4}.

\section{Conclusion and Outlook}

In conclusion, we have developed a simple model to qualitatively describe edge-state mediated absorption in s-TMD flakes and s-TMD-polymer composites. 
We show that the appropriate description must involve a three-level system, in contrast to the conventional two-band model routinely used for semiconductors.\cite{Garmire2000} 
At excitation energies near $\Delta/2$, the linear absorption of a s-TMD composite can be estimated using Eq.~(\ref{ultimate}), while the saturation fluence is given by Eq.~(\ref{ultimate2}).
The band structure parameters in Eqs.~(\ref{ultimate},\ref{ultimate2}) can be calculated\cite{Xiao2012,Kormanyos2015} or measured.\cite{Kim2016}
Our estimates of linear and saturable absorption agree, to within one order of magnitude, with existing saturable absorption measurements performed on WS$_2$ \cite{Zhang2015b} and MoSe$_2$ \cite{Woodward2015b,Luo2015} composites. 
We stress that this work does not aim at a quantitative analysis of specific samples.
For this, the following should be considered:
\begin{itemize}
 \item Due to the spin-orbit splitting in the valence band, the bandgap is different for each spin channel.
  Strictly speaking, we have {\em four} terms in the absorption $A^+_\uparrow$, $A^-_\uparrow$, $A^+_\downarrow$, $A^-_\downarrow$ instead of two $A^\pm$ considered here. 
  The non-monotonic dependence of the linear absorption on the excitation energy shown in Fig.~\ref{Fig2} becomes more complicated once we take into account of the spin-splitting.
  \item The light beam is assumed to be normal to the flakes.
The plane of incidence is therefore not well defined; consequently, our model is insensitive to s- and p-polarization.
  This is not the case in real MX$_2$ composites where flakes are randomly oriented within the host polymer matrix.
  The quantitative model should therefore include averaging not only over the azimuthal polarization angle $\theta_E$ performed here, but also  over the polar angle, as described in Ref.\cite{Paton2015} 
  \item The majority of the experimental examples of s-TMDs for ultrafast photonics exploit ultrasonic or shear assisted LPE of their bulk crystals.\cite{Howe2015} Such dispersions and composites mostly contain few-layer crystals.\cite{pssb2016howe,Woodward2015b,Zhang2015}
  In our model, we assume that the interlayer coupling is weak for the flakes produced by LPE, and any thin N-layer flake can be viewed as a stack of N monolayers, without such coupling.
  This approach works well for graphene,\cite{Sun2010} but a quantitative model for s-TMDs should address this more carefully.
  \item Our formula for saturation fluence (\ref{Phis}) is not reliable for excitation energies far from $\Delta/2$. 
  This is because at such energies, the photocarriers are excited far from the band edges and cannot be described by a thermalized distribution used here.
  In order to improve the reliability of the model a non-thermalized distribution for conduction and valence band photocarriers should be employed.
  \item To calculate the hot electron temperature, we assume that there is no energy dissipation at the time scale of the incident pump pulse duration or the electron-hole recombination process, whichever is shorter. 
  A quantitative model should include an additional term in the energy balance equation (\ref{bal1}) to take into account energy relaxation.
  \item The model neglects defects in crystals completely. These defects could result in an additional non-saturable term in Eq.~(\ref{sat2}) for the nonlinear absorption $A^\Phi$.
\end{itemize}

As an outlook we propose the following experiment to verify our model.
Our theory predicts that the edge-state optical absorption increases with the ratio of the total edge length to area
of the composite film. The crystallographic faults, impurities and other bulk defects cannot result in 
such behavior. Thus, reducing the size of flakes but keeping their mass concentration constant
we can increase the edge-state contribution to the total absorption and hence distinguish between the edge and bulk effects.

In addition to MX$_2$ flakes, our model could be applied to other hexagonal, nanostructured composites, e.g. boron nitride (\textit{h}-BN) monolayers, where the bandgap size $\Delta=3.92$~eV and the bandgap parameter $\hbar v=2.33 \mathrm{eV} \times 2.174$\AA.\cite{Ribeiro2011} 
This results in lower edge-state absorption, but the optimum excitation energy $E_\omega\approx\Delta/2$ lies in the visible region, near the wavelength of $630$nm, suggesting \textit{h}-BN may also be a suitable platform for the design of nonlinear composite-based devices in the visible spectral range.

\section{Acknowledgments}
The authors thank R. I. Woodward, R. C. T. Howe, G. Hu, and W. Belzig for fruitful discussions. EJRK and TH acknowledge funding support from the Royal Academy of Engineering (UK). 
M.T. thanks Collaborative Research Center 767 for support.

\bibliography{edgestates2.bib}

\begin{thebibliography}{46}
\expandafter\ifx\csname natexlab\endcsname\relax\def\natexlab#1{#1}\fi
\expandafter\ifx\csname bibnamefont\endcsname\relax
  \def\bibnamefont#1{#1}\fi
\expandafter\ifx\csname bibfnamefont\endcsname\relax
  \def\bibfnamefont#1{#1}\fi
\expandafter\ifx\csname citenamefont\endcsname\relax
  \def\citenamefont#1{#1}\fi
\expandafter\ifx\csname url\endcsname\relax
  \def\url#1{\texttt{#1}}\fi
\expandafter\ifx\csname urlprefix\endcsname\relax\def\urlprefix{URL }\fi
\providecommand{\bibinfo}[2]{#2}
\providecommand{\eprint}[2][]{\url{#2}}

\bibitem[{\citenamefont{Geim}(2011)}]{Geim2011}
\bibinfo{author}{\bibfnamefont{A.~K.} \bibnamefont{Geim}},
  \bibinfo{journal}{Rev. Mod. Phys.} \textbf{\bibinfo{volume}{83}},
  \bibinfo{pages}{851} (\bibinfo{year}{2011}).

\bibitem[{\citenamefont{Xu et~al.}(2013)\citenamefont{Xu, Liang, Shi, and
  Chen}}]{Xu2013}
\bibinfo{author}{\bibfnamefont{M.}~\bibnamefont{Xu}},
  \bibinfo{author}{\bibfnamefont{T.}~\bibnamefont{Liang}},
  \bibinfo{author}{\bibfnamefont{M.}~\bibnamefont{Shi}}, \bibnamefont{and}
  \bibinfo{author}{\bibfnamefont{H.}~\bibnamefont{Chen}},
  \bibinfo{journal}{Chemical Reviews} \textbf{\bibinfo{volume}{113}},
  \bibinfo{pages}{3766} (\bibinfo{year}{2013}).

\bibitem[{\citenamefont{Wang et~al.}(2012)\citenamefont{Wang, Kalantar-Zadeh,
  Kis, Coleman, and Strano}}]{Wang2012}
\bibinfo{author}{\bibfnamefont{Q.~H.} \bibnamefont{Wang}},
  \bibinfo{author}{\bibfnamefont{K.}~\bibnamefont{Kalantar-Zadeh}},
  \bibinfo{author}{\bibfnamefont{A.}~\bibnamefont{Kis}},
  \bibinfo{author}{\bibfnamefont{J.~N.} \bibnamefont{Coleman}},
  \bibnamefont{and} \bibinfo{author}{\bibfnamefont{M.~S.}
  \bibnamefont{Strano}}, \bibinfo{journal}{Nature nanotechnology}
  \textbf{\bibinfo{volume}{7}}, \bibinfo{pages}{699} (\bibinfo{year}{2012}).

\bibitem[{\citenamefont{Mak et~al.}(2010)\citenamefont{Mak, Lee, Hone, Shan,
  and Heinz}}]{Mak2010}
\bibinfo{author}{\bibfnamefont{K.~F.} \bibnamefont{Mak}},
  \bibinfo{author}{\bibfnamefont{C.}~\bibnamefont{Lee}},
  \bibinfo{author}{\bibfnamefont{J.}~\bibnamefont{Hone}},
  \bibinfo{author}{\bibfnamefont{J.}~\bibnamefont{Shan}}, \bibnamefont{and}
  \bibinfo{author}{\bibfnamefont{T.~F.} \bibnamefont{Heinz}},
  \bibinfo{journal}{Phys. Rev. Lett.} \textbf{\bibinfo{volume}{105}},
  \bibinfo{pages}{136805} (\bibinfo{year}{2010}).

\bibitem[{\citenamefont{Korn et~al.}(2011)\citenamefont{Korn, Heydrich, Hirmer,
  Schmutzler, and Sch\"uller}}]{Korn2011}
\bibinfo{author}{\bibfnamefont{T.}~\bibnamefont{Korn}},
  \bibinfo{author}{\bibfnamefont{S.}~\bibnamefont{Heydrich}},
  \bibinfo{author}{\bibfnamefont{M.}~\bibnamefont{Hirmer}},
  \bibinfo{author}{\bibfnamefont{J.}~\bibnamefont{Schmutzler}},
  \bibnamefont{and}
  \bibinfo{author}{\bibfnamefont{C.}~\bibnamefont{Sch\"uller}},
  \bibinfo{journal}{Applied Physics Letters} \textbf{\bibinfo{volume}{99}},
  \bibinfo{eid}{102109} (\bibinfo{year}{2011}).

\bibitem[{\citenamefont{Splendiani et~al.}(2010)\citenamefont{Splendiani, Sun,
  Zhang, Li, Kim, Chim, Galli, and Wang}}]{Splendiani2010}
\bibinfo{author}{\bibfnamefont{A.}~\bibnamefont{Splendiani}},
  \bibinfo{author}{\bibfnamefont{L.}~\bibnamefont{Sun}},
  \bibinfo{author}{\bibfnamefont{Y.}~\bibnamefont{Zhang}},
  \bibinfo{author}{\bibfnamefont{T.}~\bibnamefont{Li}},
  \bibinfo{author}{\bibfnamefont{J.}~\bibnamefont{Kim}},
  \bibinfo{author}{\bibfnamefont{C.-Y.} \bibnamefont{Chim}},
  \bibinfo{author}{\bibfnamefont{G.}~\bibnamefont{Galli}}, \bibnamefont{and}
  \bibinfo{author}{\bibfnamefont{F.}~\bibnamefont{Wang}},
  \bibinfo{journal}{Nano Letters} \textbf{\bibinfo{volume}{10}},
  \bibinfo{pages}{1271} (\bibinfo{year}{2010}).

\bibitem[{\citenamefont{Zhang et~al.}(2015{\natexlab{a}})\citenamefont{Zhang,
  Howe, Woodward, Kelleher, Torrisi, Hu, Popov, Taylor, and Hasan}}]{Zhang2015}
\bibinfo{author}{\bibfnamefont{M.}~\bibnamefont{Zhang}},
  \bibinfo{author}{\bibfnamefont{R.}~\bibnamefont{Howe}},
  \bibinfo{author}{\bibfnamefont{R.~I.} \bibnamefont{Woodward}},
  \bibinfo{author}{\bibfnamefont{E.~J.} \bibnamefont{Kelleher}},
  \bibinfo{author}{\bibfnamefont{F.}~\bibnamefont{Torrisi}},
  \bibinfo{author}{\bibfnamefont{G.}~\bibnamefont{Hu}},
  \bibinfo{author}{\bibfnamefont{S.}~\bibnamefont{Popov}},
  \bibinfo{author}{\bibfnamefont{J.}~\bibnamefont{Taylor}}, \bibnamefont{and}
  \bibinfo{author}{\bibfnamefont{T.}~\bibnamefont{Hasan}},
  \bibinfo{journal}{Nano Research} \textbf{\bibinfo{volume}{8}},
  \bibinfo{pages}{1522} (\bibinfo{year}{2015}{\natexlab{a}}).

\bibitem[{\citenamefont{Woodward et~al.}(2014)\citenamefont{Woodward, Kelleher,
  Howe, Hu, Torrisi, Hasan, Popov, and Taylor}}]{Woodward2014}
\bibinfo{author}{\bibfnamefont{R.~I.} \bibnamefont{Woodward}},
  \bibinfo{author}{\bibfnamefont{E.~J.~R.} \bibnamefont{Kelleher}},
  \bibinfo{author}{\bibfnamefont{R.~C.~T.} \bibnamefont{Howe}},
  \bibinfo{author}{\bibfnamefont{G.}~\bibnamefont{Hu}},
  \bibinfo{author}{\bibfnamefont{F.}~\bibnamefont{Torrisi}},
  \bibinfo{author}{\bibfnamefont{T.}~\bibnamefont{Hasan}},
  \bibinfo{author}{\bibfnamefont{S.~V.} \bibnamefont{Popov}}, \bibnamefont{and}
  \bibinfo{author}{\bibfnamefont{J.~R.} \bibnamefont{Taylor}},
  \bibinfo{journal}{Optics Express} \textbf{\bibinfo{volume}{22}},
  \bibinfo{pages}{31113} (\bibinfo{year}{2014}).

\bibitem[{\citenamefont{Zhang et~al.}(2015{\natexlab{b}})\citenamefont{Zhang,
  Hu, Hu, Howe, Chen, Zheng, and Hasan}}]{Zhang2015b}
\bibinfo{author}{\bibfnamefont{M.}~\bibnamefont{Zhang}},
  \bibinfo{author}{\bibfnamefont{G.}~\bibnamefont{Hu}},
  \bibinfo{author}{\bibfnamefont{G.}~\bibnamefont{Hu}},
  \bibinfo{author}{\bibfnamefont{R.~C.~T.} \bibnamefont{Howe}},
  \bibinfo{author}{\bibfnamefont{L.}~\bibnamefont{Chen}},
  \bibinfo{author}{\bibfnamefont{Z.}~\bibnamefont{Zheng}}, \bibnamefont{and}
  \bibinfo{author}{\bibfnamefont{T.}~\bibnamefont{Hasan}},
  \bibinfo{journal}{Scientific Reports} \textbf{\bibinfo{volume}{5}},
  \bibinfo{pages}{17482} (\bibinfo{year}{2015}{\natexlab{b}}).

\bibitem[{\citenamefont{Mao et~al.}(2015)\citenamefont{Mao, Wang, Ma, Han,
  Jiang, Gan, Hua, Zhang, Mei, and Zhao}}]{Mao2015}
\bibinfo{author}{\bibfnamefont{D.}~\bibnamefont{Mao}},
  \bibinfo{author}{\bibfnamefont{Y.}~\bibnamefont{Wang}},
  \bibinfo{author}{\bibfnamefont{C.}~\bibnamefont{Ma}},
  \bibinfo{author}{\bibfnamefont{L.}~\bibnamefont{Han}},
  \bibinfo{author}{\bibfnamefont{B.}~\bibnamefont{Jiang}},
  \bibinfo{author}{\bibfnamefont{X.}~\bibnamefont{Gan}},
  \bibinfo{author}{\bibfnamefont{S.}~\bibnamefont{Hua}},
  \bibinfo{author}{\bibfnamefont{W.}~\bibnamefont{Zhang}},
  \bibinfo{author}{\bibfnamefont{T.}~\bibnamefont{Mei}}, \bibnamefont{and}
  \bibinfo{author}{\bibfnamefont{J.}~\bibnamefont{Zhao}},
  \bibinfo{journal}{Scientific reports} \textbf{\bibinfo{volume}{5}},
  \bibinfo{pages}{7965} (\bibinfo{year}{2015}).

\bibitem[{\citenamefont{Woodward
  et~al.}(2015{\natexlab{a}})\citenamefont{Woodward, Howe, Runcorn, Hu,
  Torrisi, Kelleher, and Hasan}}]{Woodward2015b}
\bibinfo{author}{\bibfnamefont{R.~I.} \bibnamefont{Woodward}},
  \bibinfo{author}{\bibfnamefont{R.~C.~T.} \bibnamefont{Howe}},
  \bibinfo{author}{\bibfnamefont{T.~H.} \bibnamefont{Runcorn}},
  \bibinfo{author}{\bibfnamefont{G.}~\bibnamefont{Hu}},
  \bibinfo{author}{\bibfnamefont{F.}~\bibnamefont{Torrisi}},
  \bibinfo{author}{\bibfnamefont{E.~J.~R.} \bibnamefont{Kelleher}},
  \bibnamefont{and} \bibinfo{author}{\bibfnamefont{T.}~\bibnamefont{Hasan}},
  \bibinfo{journal}{Opt. Express} \textbf{\bibinfo{volume}{23}},
  \bibinfo{pages}{20051} (\bibinfo{year}{2015}{\natexlab{a}}).

\bibitem[{\citenamefont{Luo et~al.}(2015)\citenamefont{Luo, Li, Zhong, Huang,
  Wan, Peng, and Weng}}]{Luo2015}
\bibinfo{author}{\bibfnamefont{Z.}~\bibnamefont{Luo}},
  \bibinfo{author}{\bibfnamefont{Y.}~\bibnamefont{Li}},
  \bibinfo{author}{\bibfnamefont{M.}~\bibnamefont{Zhong}},
  \bibinfo{author}{\bibfnamefont{Y.}~\bibnamefont{Huang}},
  \bibinfo{author}{\bibfnamefont{X.}~\bibnamefont{Wan}},
  \bibinfo{author}{\bibfnamefont{J.}~\bibnamefont{Peng}}, \bibnamefont{and}
  \bibinfo{author}{\bibfnamefont{J.}~\bibnamefont{Weng}},
  \bibinfo{journal}{Photonics Research} \textbf{\bibinfo{volume}{3}},
  \bibinfo{pages}{A79} (\bibinfo{year}{2015}).

\bibitem[{\citenamefont{Zhang et~al.}(2014)\citenamefont{Zhang, Chang, Zhou,
  Cui, Yan, Liu, Schmitt, Lee, Moore, Chen et~al.}}]{Zhang2014}
\bibinfo{author}{\bibfnamefont{Y.}~\bibnamefont{Zhang}},
  \bibinfo{author}{\bibfnamefont{T.-R.} \bibnamefont{Chang}},
  \bibinfo{author}{\bibfnamefont{B.}~\bibnamefont{Zhou}},
  \bibinfo{author}{\bibfnamefont{Y.-T.} \bibnamefont{Cui}},
  \bibinfo{author}{\bibfnamefont{H.}~\bibnamefont{Yan}},
  \bibinfo{author}{\bibfnamefont{Z.}~\bibnamefont{Liu}},
  \bibinfo{author}{\bibfnamefont{F.}~\bibnamefont{Schmitt}},
  \bibinfo{author}{\bibfnamefont{J.}~\bibnamefont{Lee}},
  \bibinfo{author}{\bibfnamefont{R.}~\bibnamefont{Moore}},
  \bibinfo{author}{\bibfnamefont{Y.}~\bibnamefont{Chen}}, \bibnamefont{et~al.},
  \bibinfo{journal}{Nature nanotechnology} \textbf{\bibinfo{volume}{9}},
  \bibinfo{pages}{111} (\bibinfo{year}{2014}).

\bibitem[{\citenamefont{Wang et~al.}(2014)\citenamefont{Wang, Yu, Zhang, Wang,
  Zhao, Chen, Mei, and Wang}}]{Wang2014}
\bibinfo{author}{\bibfnamefont{S.}~\bibnamefont{Wang}},
  \bibinfo{author}{\bibfnamefont{H.}~\bibnamefont{Yu}},
  \bibinfo{author}{\bibfnamefont{H.}~\bibnamefont{Zhang}},
  \bibinfo{author}{\bibfnamefont{A.}~\bibnamefont{Wang}},
  \bibinfo{author}{\bibfnamefont{M.}~\bibnamefont{Zhao}},
  \bibinfo{author}{\bibfnamefont{Y.}~\bibnamefont{Chen}},
  \bibinfo{author}{\bibfnamefont{L.}~\bibnamefont{Mei}}, \bibnamefont{and}
  \bibinfo{author}{\bibfnamefont{J.}~\bibnamefont{Wang}},
  \bibinfo{journal}{Advanced materials} \textbf{\bibinfo{volume}{26}},
  \bibinfo{pages}{3538} (\bibinfo{year}{2014}).

\bibitem[{\citenamefont{Woodward
  et~al.}(2015{\natexlab{b}})\citenamefont{Woodward, Howe, Hu, Torrisi, Zhang,
  Hasan, and Kelleher}}]{Woodward2015}
\bibinfo{author}{\bibfnamefont{R.~I.} \bibnamefont{Woodward}},
  \bibinfo{author}{\bibfnamefont{R.~C.~T.} \bibnamefont{Howe}},
  \bibinfo{author}{\bibfnamefont{G.}~\bibnamefont{Hu}},
  \bibinfo{author}{\bibfnamefont{F.}~\bibnamefont{Torrisi}},
  \bibinfo{author}{\bibfnamefont{M.}~\bibnamefont{Zhang}},
  \bibinfo{author}{\bibfnamefont{T.}~\bibnamefont{Hasan}}, \bibnamefont{and}
  \bibinfo{author}{\bibfnamefont{E.~J.~R.} \bibnamefont{Kelleher}},
  \bibinfo{journal}{Photon. Res.} \textbf{\bibinfo{volume}{3}},
  \bibinfo{pages}{A30} (\bibinfo{year}{2015}{\natexlab{b}}).

\bibitem[{\citenamefont{Howe et~al.}(2016)\citenamefont{Howe, Woodward, Hu,
  Yang, Kelleher, and Hasan}}]{pssb2016howe}
\bibinfo{author}{\bibfnamefont{R.~C.~T.} \bibnamefont{Howe}},
  \bibinfo{author}{\bibfnamefont{R.~I.} \bibnamefont{Woodward}},
  \bibinfo{author}{\bibfnamefont{G.}~\bibnamefont{Hu}},
  \bibinfo{author}{\bibfnamefont{Z.}~\bibnamefont{Yang}},
  \bibinfo{author}{\bibfnamefont{E.~J.~R.} \bibnamefont{Kelleher}},
  \bibnamefont{and} \bibinfo{author}{\bibfnamefont{T.}~\bibnamefont{Hasan}},
  \bibinfo{journal}{physica status solidi (b)} \textbf{\bibinfo{volume}{253}},
  \bibinfo{pages}{911} (\bibinfo{year}{2016}).

\bibitem[{\citenamefont{Roxlo et~al.}(1987)\citenamefont{Roxlo, Chianelli,
  Deckman, Ruppert, and Wong}}]{Roxlo1987}
\bibinfo{author}{\bibfnamefont{C.~B.} \bibnamefont{Roxlo}},
  \bibinfo{author}{\bibfnamefont{R.~R.} \bibnamefont{Chianelli}},
  \bibinfo{author}{\bibfnamefont{H.~W.} \bibnamefont{Deckman}},
  \bibinfo{author}{\bibfnamefont{A.~F.} \bibnamefont{Ruppert}},
  \bibnamefont{and} \bibinfo{author}{\bibfnamefont{P.~P.} \bibnamefont{Wong}},
  \bibinfo{journal}{Journal of Vacuum Science \& Technology A}
  \textbf{\bibinfo{volume}{5}}, \bibinfo{pages}{555} (\bibinfo{year}{1987}).

\bibitem[{\citenamefont{P\'eterfalvi et~al.}(2015)\citenamefont{P\'eterfalvi,
  Korm\'anyos, and Burkard}}]{Peterfalvi2015}
\bibinfo{author}{\bibfnamefont{C.~G.} \bibnamefont{P\'eterfalvi}},
  \bibinfo{author}{\bibfnamefont{A.}~\bibnamefont{Korm\'anyos}},
  \bibnamefont{and} \bibinfo{author}{\bibfnamefont{G.}~\bibnamefont{Burkard}},
  \bibinfo{journal}{Phys. Rev. B} \textbf{\bibinfo{volume}{92}},
  \bibinfo{pages}{245443} (\bibinfo{year}{2015}).

\bibitem[{\citenamefont{Brey and Fertig}(2006)}]{Brey2006}
\bibinfo{author}{\bibfnamefont{L.}~\bibnamefont{Brey}} \bibnamefont{and}
  \bibinfo{author}{\bibfnamefont{H.~A.} \bibnamefont{Fertig}},
  \bibinfo{journal}{Phys. Rev. B} \textbf{\bibinfo{volume}{73}},
  \bibinfo{pages}{235411} (\bibinfo{year}{2006}).

\bibitem[{\citenamefont{Akhmerov and Beenakker}(2008)}]{Akhmerov2008}
\bibinfo{author}{\bibfnamefont{A.~R.} \bibnamefont{Akhmerov}} \bibnamefont{and}
  \bibinfo{author}{\bibfnamefont{C.~W.~J.} \bibnamefont{Beenakker}},
  \bibinfo{journal}{Phys. Rev. B} \textbf{\bibinfo{volume}{77}},
  \bibinfo{pages}{085423} (\bibinfo{year}{2008}).

\bibitem[{\citenamefont{Lado et~al.}(2015)\citenamefont{Lado,
  Garci�a-Marti�nez, and Fernandez-Rossier}}]{Lado2015}
\bibinfo{author}{\bibfnamefont{J.}~\bibnamefont{Lado}},
  \bibinfo{author}{\bibfnamefont{N.}~\bibnamefont{Garci�a-Marti�nez}},
  \bibnamefont{and}
  \bibinfo{author}{\bibfnamefont{J.}~\bibnamefont{Fernandez-Rossier}},
  \bibinfo{journal}{Synthetic Metals} \textbf{\bibinfo{volume}{210}},
  \bibinfo{pages}{56} (\bibinfo{year}{2015}).

\bibitem[{\citenamefont{Pavlovi\ifmmode~\acute{c}\else \'{c}\fi{} and
  Peeters}(2015)}]{Pavlovic2015}
\bibinfo{author}{\bibfnamefont{S.}~\bibnamefont{Pavlovi\ifmmode~\acute{c}\else
  \'{c}\fi{}}} \bibnamefont{and} \bibinfo{author}{\bibfnamefont{F.~M.}
  \bibnamefont{Peeters}}, \bibinfo{journal}{Phys. Rev. B}
  \textbf{\bibinfo{volume}{91}}, \bibinfo{pages}{155410}
  (\bibinfo{year}{2015}).

\bibitem[{\citenamefont{Segarra et~al.}(2016)\citenamefont{Segarra, Planelles,
  and Ulloa}}]{Segarra2015}
\bibinfo{author}{\bibfnamefont{C.}~\bibnamefont{Segarra}},
  \bibinfo{author}{\bibfnamefont{J.}~\bibnamefont{Planelles}},
  \bibnamefont{and} \bibinfo{author}{\bibfnamefont{S.~E.} \bibnamefont{Ulloa}},
  \bibinfo{journal}{Phys. Rev. B} \textbf{\bibinfo{volume}{93}},
  \bibinfo{pages}{085312} (\bibinfo{year}{2016}).

\bibitem[{\citenamefont{Bollinger et~al.}(2001)\citenamefont{Bollinger,
  Lauritsen, Jacobsen, N\o{}rskov, Helveg, and Besenbacher}}]{PRL2001bollinger}
\bibinfo{author}{\bibfnamefont{M.~V.} \bibnamefont{Bollinger}},
  \bibinfo{author}{\bibfnamefont{J.~V.} \bibnamefont{Lauritsen}},
  \bibinfo{author}{\bibfnamefont{K.~W.} \bibnamefont{Jacobsen}},
  \bibinfo{author}{\bibfnamefont{J.~K.} \bibnamefont{N\o{}rskov}},
  \bibinfo{author}{\bibfnamefont{S.}~\bibnamefont{Helveg}}, \bibnamefont{and}
  \bibinfo{author}{\bibfnamefont{F.}~\bibnamefont{Besenbacher}},
  \bibinfo{journal}{Phys. Rev. Lett.} \textbf{\bibinfo{volume}{87}},
  \bibinfo{pages}{196803} (\bibinfo{year}{2001}).

\bibitem[{\citenamefont{Bollinger et~al.}(2003)\citenamefont{Bollinger,
  Jacobsen, and Norskov}}]{Bollinger2003}
\bibinfo{author}{\bibfnamefont{M.~V.} \bibnamefont{Bollinger}},
  \bibinfo{author}{\bibfnamefont{K.~W.} \bibnamefont{Jacobsen}},
  \bibnamefont{and} \bibinfo{author}{\bibfnamefont{J.~K.}
  \bibnamefont{Norskov}}, \bibinfo{journal}{Phys. Rev. B}
  \textbf{\bibinfo{volume}{67}}, \bibinfo{pages}{085410}
  (\bibinfo{year}{2003}).

\bibitem[{\citenamefont{Li et~al.}(2008)\citenamefont{Li, Zhou, Zhang, and
  Chen}}]{JAMS2008li}
\bibinfo{author}{\bibfnamefont{Y.}~\bibnamefont{Li}},
  \bibinfo{author}{\bibfnamefont{Z.}~\bibnamefont{Zhou}},
  \bibinfo{author}{\bibfnamefont{S.}~\bibnamefont{Zhang}}, \bibnamefont{and}
  \bibinfo{author}{\bibfnamefont{Z.}~\bibnamefont{Chen}},
  \bibinfo{journal}{Journal of the American Chemical Society}
  \textbf{\bibinfo{volume}{130}}, \bibinfo{pages}{16739}
  (\bibinfo{year}{2008}).

\bibitem[{\citenamefont{Vojvodic et~al.}(2009)\citenamefont{Vojvodic,
  Hinnemann, and N\o{}rskov}}]{Vojvodic2009}
\bibinfo{author}{\bibfnamefont{A.}~\bibnamefont{Vojvodic}},
  \bibinfo{author}{\bibfnamefont{B.}~\bibnamefont{Hinnemann}},
  \bibnamefont{and} \bibinfo{author}{\bibfnamefont{J.~K.}
  \bibnamefont{N\o{}rskov}}, \bibinfo{journal}{Phys. Rev. B}
  \textbf{\bibinfo{volume}{80}}, \bibinfo{pages}{125416}
  (\bibinfo{year}{2009}).

\bibitem[{\citenamefont{Kou et~al.}(2012)\citenamefont{Kou, Tang, Zhang, Heine,
  Chen, and Frauenheim}}]{JPCL2012kou}
\bibinfo{author}{\bibfnamefont{L.}~\bibnamefont{Kou}},
  \bibinfo{author}{\bibfnamefont{C.}~\bibnamefont{Tang}},
  \bibinfo{author}{\bibfnamefont{Y.}~\bibnamefont{Zhang}},
  \bibinfo{author}{\bibfnamefont{T.}~\bibnamefont{Heine}},
  \bibinfo{author}{\bibfnamefont{C.}~\bibnamefont{Chen}}, \bibnamefont{and}
  \bibinfo{author}{\bibfnamefont{T.}~\bibnamefont{Frauenheim}},
  \bibinfo{journal}{The Journal of Physical Chemistry Letters}
  \textbf{\bibinfo{volume}{3}}, \bibinfo{pages}{2934} (\bibinfo{year}{2012}).

\bibitem[{\citenamefont{Erdogan et~al.}(2012)\citenamefont{Erdogan, Popov,
  Enyashin, and Seifert}}]{Erdogan2012}
\bibinfo{author}{\bibfnamefont{E.}~\bibnamefont{Erdogan}},
  \bibinfo{author}{\bibfnamefont{I.}~\bibnamefont{Popov}},
  \bibinfo{author}{\bibfnamefont{A.}~\bibnamefont{Enyashin}}, \bibnamefont{and}
  \bibinfo{author}{\bibfnamefont{G.}~\bibnamefont{Seifert}},
  \bibinfo{journal}{The European Physical Journal B}
  \textbf{\bibinfo{volume}{85}}, \bibinfo{eid}{33} (\bibinfo{year}{2012}).

\bibitem[{\citenamefont{Xu et~al.}(2014)\citenamefont{Xu, Wang, Yan, and
  Qi}}]{PRB2014xu}
\bibinfo{author}{\bibfnamefont{G.}~\bibnamefont{Xu}},
  \bibinfo{author}{\bibfnamefont{J.}~\bibnamefont{Wang}},
  \bibinfo{author}{\bibfnamefont{B.}~\bibnamefont{Yan}}, \bibnamefont{and}
  \bibinfo{author}{\bibfnamefont{X.-L.} \bibnamefont{Qi}},
  \bibinfo{journal}{Phys. Rev. B} \textbf{\bibinfo{volume}{90}},
  \bibinfo{pages}{100505} (\bibinfo{year}{2014}).

\bibitem[{\citenamefont{Pan and Zhang}(2012)}]{RSC2012pan}
\bibinfo{author}{\bibfnamefont{H.}~\bibnamefont{Pan}} \bibnamefont{and}
  \bibinfo{author}{\bibfnamefont{Y.-W.} \bibnamefont{Zhang}},
  \bibinfo{journal}{J. Mater. Chem.} \textbf{\bibinfo{volume}{22}},
  \bibinfo{pages}{7280} (\bibinfo{year}{2012}).

\bibitem[{\citenamefont{Xiao et~al.}(2012)\citenamefont{Xiao, Liu, Feng, Xu,
  and Yao}}]{Xiao2012}
\bibinfo{author}{\bibfnamefont{D.}~\bibnamefont{Xiao}},
  \bibinfo{author}{\bibfnamefont{G.-B.} \bibnamefont{Liu}},
  \bibinfo{author}{\bibfnamefont{W.}~\bibnamefont{Feng}},
  \bibinfo{author}{\bibfnamefont{X.}~\bibnamefont{Xu}}, \bibnamefont{and}
  \bibinfo{author}{\bibfnamefont{W.}~\bibnamefont{Yao}},
  \bibinfo{journal}{Phys. Rev. Lett.} \textbf{\bibinfo{volume}{108}},
  \bibinfo{pages}{196802} (\bibinfo{year}{2012}).

\bibitem[{\citenamefont{Berry and Mondragon}(1987)}]{Berry1987}
\bibinfo{author}{\bibfnamefont{M.~V.} \bibnamefont{Berry}} \bibnamefont{and}
  \bibinfo{author}{\bibfnamefont{R.~J.} \bibnamefont{Mondragon}},
  \bibinfo{journal}{Proceedings of the Royal Society of London A: Mathematical,
  Physical and Engineering Sciences} \textbf{\bibinfo{volume}{412}},
  \bibinfo{pages}{53} (\bibinfo{year}{1987}).

\bibitem[{\citenamefont{Garmire}(2000)}]{Garmire2000}
\bibinfo{author}{\bibfnamefont{E.}~\bibnamefont{Garmire}},
  \bibinfo{journal}{Selected Topics in Quantum Electronics, IEEE Journal of}
  \textbf{\bibinfo{volume}{6}}, \bibinfo{pages}{1094} (\bibinfo{year}{2000}).

\bibitem[{\citenamefont{McCann�}(2012)}]{McCann2012}
\bibinfo{author}{\bibfnamefont{E.}~\bibnamefont{McCann�}},
  \emph{\bibinfo{title}{Graphene Nanoelectronics: Metrology, Synthesis,
  Properties and Applications, chapter 8}} (\bibinfo{publisher}{Springer-Verlag
  Berlin Heidelberg}, \bibinfo{year}{2012}).

\bibitem[{\citenamefont{Hasan and Kane}(2010)}]{Hasan2010}
\bibinfo{author}{\bibfnamefont{M.~Z.} \bibnamefont{Hasan}} \bibnamefont{and}
  \bibinfo{author}{\bibfnamefont{C.~L.} \bibnamefont{Kane}},
  \bibinfo{journal}{Rev. Mod. Phys.} \textbf{\bibinfo{volume}{82}},
  \bibinfo{pages}{3045} (\bibinfo{year}{2010}).

\bibitem[{\citenamefont{Trushin et~al.}(2015)\citenamefont{Trushin, Grupp,
  Soavi, Budweg, De~Fazio, Sassi, Lombardo, Ferrari, Belzig, Leitenstorfer
  et~al.}}]{Trushin2015}
\bibinfo{author}{\bibfnamefont{M.}~\bibnamefont{Trushin}},
  \bibinfo{author}{\bibfnamefont{A.}~\bibnamefont{Grupp}},
  \bibinfo{author}{\bibfnamefont{G.}~\bibnamefont{Soavi}},
  \bibinfo{author}{\bibfnamefont{A.}~\bibnamefont{Budweg}},
  \bibinfo{author}{\bibfnamefont{D.}~\bibnamefont{De~Fazio}},
  \bibinfo{author}{\bibfnamefont{U.}~\bibnamefont{Sassi}},
  \bibinfo{author}{\bibfnamefont{A.}~\bibnamefont{Lombardo}},
  \bibinfo{author}{\bibfnamefont{A.~C.} \bibnamefont{Ferrari}},
  \bibinfo{author}{\bibfnamefont{W.}~\bibnamefont{Belzig}},
  \bibinfo{author}{\bibfnamefont{A.}~\bibnamefont{Leitenstorfer}},
  \bibnamefont{et~al.}, \bibinfo{journal}{Phys. Rev. B}
  \textbf{\bibinfo{volume}{92}}, \bibinfo{pages}{165429}
  (\bibinfo{year}{2015}).

\bibitem[{\citenamefont{Trushin and Schliemann}(2011)}]{Trushin2011}
\bibinfo{author}{\bibfnamefont{M.}~\bibnamefont{Trushin}} \bibnamefont{and}
  \bibinfo{author}{\bibfnamefont{J.}~\bibnamefont{Schliemann}},
  \bibinfo{journal}{EPL (Europhysics Letters)} \textbf{\bibinfo{volume}{96}},
  \bibinfo{pages}{37006} (\bibinfo{year}{2011}).

\bibitem[{\citenamefont{Nelson}(2004)}]{Nelson2004}
\bibinfo{author}{\bibfnamefont{J.}~\bibnamefont{Nelson}},
  \emph{\bibinfo{title}{The Physics of Solar Cells}}
  (\bibinfo{publisher}{Imperial College Press, UK}, \bibinfo{year}{2004}).

\bibitem[{\citenamefont{Kim et~al.}(2016)\citenamefont{Kim, Rhim, Kim, Kim, and
  Park}}]{Kim2016}
\bibinfo{author}{\bibfnamefont{B.~S.} \bibnamefont{Kim}},
  \bibinfo{author}{\bibfnamefont{J.-W.} \bibnamefont{Rhim}},
  \bibinfo{author}{\bibfnamefont{B.}~\bibnamefont{Kim}},
  \bibinfo{author}{\bibfnamefont{C.}~\bibnamefont{Kim}}, \bibnamefont{and}
  \bibinfo{author}{\bibfnamefont{S.~R.} \bibnamefont{Park}},
  \bibinfo{journal}{arXiv:1601.01418 [cond-mat.mes-hall]}
  (\bibinfo{year}{2016}).

\bibitem[{\citenamefont{Wang et~al.}(2015)\citenamefont{Wang, Zhang, and
  Rana}}]{Wang2015}
\bibinfo{author}{\bibfnamefont{H.}~\bibnamefont{Wang}},
  \bibinfo{author}{\bibfnamefont{C.}~\bibnamefont{Zhang}}, \bibnamefont{and}
  \bibinfo{author}{\bibfnamefont{F.}~\bibnamefont{Rana}},
  \bibinfo{journal}{Nano Letters} \textbf{\bibinfo{volume}{15}},
  \bibinfo{pages}{8204} (\bibinfo{year}{2015}).

\bibitem[{\citenamefont{Korm\'{a}nyos et~al.}(2015)\citenamefont{Korm\'{a}nyos,
  Burkard, Gmitra, Fabian, Z\'{o}lyomi, Drummond, and Fal'ko}}]{Kormanyos2015}
\bibinfo{author}{\bibfnamefont{A.}~\bibnamefont{Korm\'{a}nyos}},
  \bibinfo{author}{\bibfnamefont{G.}~\bibnamefont{Burkard}},
  \bibinfo{author}{\bibfnamefont{M.}~\bibnamefont{Gmitra}},
  \bibinfo{author}{\bibfnamefont{J.}~\bibnamefont{Fabian}},
  \bibinfo{author}{\bibfnamefont{V.}~\bibnamefont{Z\'{o}lyomi}},
  \bibinfo{author}{\bibfnamefont{N.~D.} \bibnamefont{Drummond}},
  \bibnamefont{and} \bibinfo{author}{\bibfnamefont{V.}~\bibnamefont{Fal'ko}},
  \bibinfo{journal}{2D Materials} \textbf{\bibinfo{volume}{2}},
  \bibinfo{pages}{022001} (\bibinfo{year}{2015}).

\bibitem[{\citenamefont{Paton and Coleman}(2015)}]{Paton2015}
\bibinfo{author}{\bibfnamefont{K.~R.} \bibnamefont{Paton}} \bibnamefont{and}
  \bibinfo{author}{\bibfnamefont{J.~N.} \bibnamefont{Coleman}},
  \bibinfo{journal}{arXiv:1511.04410 [cond-mat.mes-hall]}
  (\bibinfo{year}{2015}).

\bibitem[{\citenamefont{Howe et~al.}(2015)\citenamefont{Howe, Hu, Yang, and
  Hasan}}]{Howe2015}
\bibinfo{author}{\bibfnamefont{R.~C.~T.} \bibnamefont{Howe}},
  \bibinfo{author}{\bibfnamefont{G.}~\bibnamefont{Hu}},
  \bibinfo{author}{\bibfnamefont{Z.}~\bibnamefont{Yang}}, \bibnamefont{and}
  \bibinfo{author}{\bibfnamefont{T.}~\bibnamefont{Hasan}}, in
  \emph{\bibinfo{booktitle}{Proc. SPIE 9553, Low-Dimensional Materials and
  Devices}} (\bibinfo{year}{2015}), vol. \bibinfo{volume}{9553}, p.
  \bibinfo{pages}{95530R}.

\bibitem[{\citenamefont{Sun et~al.}(2010)\citenamefont{Sun, Hasan, Torrisi,
  Popa, Privitera, Wang, Bonaccorso, Basko, and Ferrari}}]{Sun2010}
\bibinfo{author}{\bibfnamefont{Z.}~\bibnamefont{Sun}},
  \bibinfo{author}{\bibfnamefont{T.}~\bibnamefont{Hasan}},
  \bibinfo{author}{\bibfnamefont{F.}~\bibnamefont{Torrisi}},
  \bibinfo{author}{\bibfnamefont{D.}~\bibnamefont{Popa}},
  \bibinfo{author}{\bibfnamefont{G.}~\bibnamefont{Privitera}},
  \bibinfo{author}{\bibfnamefont{F.}~\bibnamefont{Wang}},
  \bibinfo{author}{\bibfnamefont{F.}~\bibnamefont{Bonaccorso}},
  \bibinfo{author}{\bibfnamefont{D.~M.} \bibnamefont{Basko}}, \bibnamefont{and}
  \bibinfo{author}{\bibfnamefont{A.~C.} \bibnamefont{Ferrari}},
  \bibinfo{journal}{ACS Nano} \textbf{\bibinfo{volume}{4}},
  \bibinfo{pages}{803} (\bibinfo{year}{2010}).

\bibitem[{\citenamefont{Ribeiro and Peres}(2011)}]{Ribeiro2011}
\bibinfo{author}{\bibfnamefont{R.~M.} \bibnamefont{Ribeiro}} \bibnamefont{and}
  \bibinfo{author}{\bibfnamefont{N.~M.~R.} \bibnamefont{Peres}},
  \bibinfo{journal}{Phys. Rev. B} \textbf{\bibinfo{volume}{83}},
  \bibinfo{pages}{235312} (\bibinfo{year}{2011}).

\end{thebibliography}

\end{document}